\documentclass[conference,10pt,letterpaper]{IEEEtran}
\IEEEoverridecommandlockouts

\usepackage{mathtools}
\usepackage{amsmath}
\usepackage{etoolbox}
\usepackage{booktabs}
\usepackage{cite}
\usepackage{amsfonts}
\usepackage{xurl}
\usepackage{siunitx}
\usepackage[linesnumbered,titlenumbered,ruled,vlined,noend]{algorithm2e}
\SetAlCapFnt{\normalsize}
\SetAlCapNameFnt{\normalfont\normalsize}
\SetAlCapSty{textbf}
\SetAlCapNameSty{textnormal}
\SetAlgoCaptionSeparator{:}
\makeatletter
\patchcmd\algocf@Vline{\vrule}{\vrule \kern-0.4pt}{}{}
\patchcmd\algocf@Vsline{\vrule}{\vrule \kern-0.4pt}{}{}
\makeatother

\usepackage{enumerate}
\usepackage{enumitem}
\newlist{Properties}{enumerate}{1}
\setlist[Properties]{label=Property \arabic*.,font=\textbf,itemindent=*,leftmargin=0pt}
\usepackage{color}
\usepackage{makecell}
\usepackage[normalem]{ulem}
\usepackage{tikz}
\usepackage{graphicx}
\usetikzlibrary{positioning,arrows.meta,fit,calc,shapes.geometric}

\usepackage{comment}

\usepackage{csquotes}

\usepackage{float}
\usepackage{placeins}
\usepackage{caption}
\SetAlgoSkip{ObliSCAlgoSkip}
\makeatletter
\newif\ifObliSCCompactFloat
\newcommand{\ObliSCTrackFloat}[1]{%
  \ifnum\count#1<96\relax
    \ObliSCCompactFloattrue
  \else
    \ObliSCCompactFloatfalse
  \fi}
\newcommand{\ObliSCBelowFloatSkip}[1]{%
  \ifObliSCCompactFloat
    \vskip\glueexpr#1/2\relax
  \else
    \vskip#1%
  \fi}
\newcommand{\ObliSCFloatPatchError}{%
  \PackageError{ObliSC}{Float-spacing patch failed}{Check the LaTeX output routine.}}
\pretocmd{\@comflelt}{\ObliSCTrackFloat{#1}}{}{\ObliSCFloatPatchError}
\pretocmd{\@comdblflelt}{\ObliSCTrackFloat{#1}}{}{\ObliSCFloatPatchError}
\patchcmd{\@cflt}{\vskip\textfloatsep}
  {\ObliSCBelowFloatSkip{\textfloatsep}}{}{\ObliSCFloatPatchError}
\patchcmd{\@combinedblfloats}{\vskip\dbltextfloatsep}
  {\ObliSCBelowFloatSkip{\dbltextfloatsep}}{}{\ObliSCFloatPatchError}
\patchcmd{\@addtocurcol}
  {\box\@currbox\penalty\interlinepenalty\vskip\intextsep}
  {\ObliSCTrackFloat{\@currbox}\box\@currbox\penalty\interlinepenalty
   \ObliSCBelowFloatSkip{\intextsep}}{}{\ObliSCFloatPatchError}
\makeatother

\usepackage{soul}
\usepackage{pifont} %
\usepackage{multirow}
\usepackage{footnote}
\usepackage{amsthm}
\usepackage{balance}

\makesavenoteenv{figure}
\usepackage{xcolor}
\usepackage{colortbl}
\definecolor{MyLightBlue}{RGB}{155, 210, 255}
\definecolor{AccessPatternShade}{RGB}{235,235,235}

\providecommand{\Description}[1]{}

\setlist{nosep, topsep=2pt, partopsep=1pt}

\SetKwRepeat{Do}{do}{while}%

\let\oldnl\nl%
\newcommand{\nonl}{\renewcommand{\nl}{\let\nl\oldnl}}%

\newcommand{\stab}{\rule{0pt}{8pt}\\[-2ex]}
\newcommand{\eg}{{\it e.g.}, }
\newcommand{\ie}{{\it i.e.}, }
\newcommand{\wrt}{{\it w.r.t} }
\newcommand{\resp}{{\it resp.} }

\definecolor{darkgreen}{RGB}{0,100,0}

\newcommand{\name}{{{\mathsf{ObliSC}}}}
\newcommand{\SP}{{{\mathsf{SP}}}}
\newcommand{\access}{\mathsf{Access}}
\newcommand{\FO}{\mathsf{FO}}

\newcommand{\ann}{\mathsf{A}k\mathsf{NN}}

\newtheorem{definition}{Definition}
\newtheorem{example}{Example}

\newtheorem{theorem}{Theorem}
\newtheorem{lemma}{Lemma}

\usepackage[hidelinks]{hyperref}

\begin{document}

\title{Efficient Privacy-Preserving High-Dimensional Approximate Nearest Neighbor Search}

\newcommand{\ppscauthorentry}[4]{%
\begin{minipage}[t]{0.31\textwidth}
\centering
#1\par
\vspace{0.25ex}
{\normalsize #2\\#3\\#4\par}
\end{minipage}%
}

\author{
\IEEEauthorblockN{
Lyu Xu\IEEEauthorrefmark{1},
Guozhong Li\IEEEauthorrefmark{2},
and Qichen Wang\IEEEauthorrefmark{1}
}
\IEEEauthorblockA{
\IEEEauthorrefmark{1}Nanyang Technological University, Singapore
}
\IEEEauthorblockA{
\IEEEauthorrefmark{2}King Abdullah University of Science and Technology, Thuwal, Saudi Arabia
}
\IEEEauthorblockA{
\IEEEauthorrefmark{1}\{lyu.xu,qichen.wang\}@ntu.edu.sg,
\IEEEauthorrefmark{2}guozhong.li@kaust.edu.sa
}
}

\maketitle

\begin{abstract}
In the AI era, approximate nearest neighbor (ANN) search on vector databases is critical as it supports similarity retrieval over high-dimensional vectors, which is essential for many large language model (LLM) applications. To reduce costs and improve scalability, vector databases are often outsourced to cloud servers, raising privacy concerns for users whose query vectors contain sensitive information. In this work, we study the problem of asymmetric privacy-preserving ANN search, where the vector database is public but user queries must remain private from an untrusted cloud server. To solve this problem, we propose $\mathsf{ObliSC}$, an efficient privacy-preserving search framework that protects query privacy against access-pattern leakage. $\mathsf{ObliSC}$ consists of (i) a search base that retrieves ANNs with a strong privacy guarantee of full obliviousness; and (ii) parallel optimization strategies that accelerate candidate retrieval and refinement while preserving full obliviousness and retrieval quality. Experiments on six standard ANN benchmarks show that $\mathsf{ObliSC}$ not only achieves efficiency and result quality competitive with baseline methods but also maintains a lightweight index compared with the evaluated privacy-preserving approaches.
\end{abstract}
\begin{IEEEkeywords}
Approximate nearest neighbor search, query privacy, access pattern protection, obliviousness.
\end{IEEEkeywords}

\section{Introduction}\label{sec:intro}
Vector databases support AI applications that retrieve objects by semantic similarity. In retrieval-augmented generation (RAG), for example, a dense retriever embeds a question and searches a passage index for context that a language model can use to generate an answer~\cite{lewis2020retrieval}. Approximate nearest neighbor (ANN) search is a basic operation for such retrieval: it finds vectors close to the query without comparing it with every item in a large collection~\cite{malkov2020hnsw, peng2023efficient, wei2025suco}.

Nowadays, the cloud service is popular since the computational resources of individual users are limited. In this scenario, the query and the collection may have different privacy requirements. Consider a publisher that hosts a vector index of its technical manuals in its own cloud. The publisher already knows the manuals and can inspect their vectors in plaintext. A customer, however, may search the manuals with an embedding of an unpublished incident report. Even revealing which manuals the customer consulted could identify the product or fault under investigation. As there are increasing concerns about disclosing personal information~\cite{norberg2007privacy}, privacy rights and data protection~\cite{nyst2017right,GDPR2016}, and the design of privacy-preserving services~\cite{apple2021}, it is critical to protect what users reveal when they interact with a service provider (SP). Hence, in this paper, we study the corresponding \emph{asymmetric} ANN search problem: the database and index are public to SP, whereas the user's query must remain private from SP.

Prior work such as Preco~\cite{DBLP:conf/sp/Servan-Schreiber22} protects the user's query while bounding what the client learns about the database. However, Preco requires the user to query two servers and relies on them not pooling their respective query shares. In contrast, we focus on a single-server design for asymmetric ANN search that does not require a second, non-colluding server.

\begin{table*}[t]
\centering
\footnotesize
\caption{Summary of representative privacy-preserving ANN approaches. 
Quality and cost are qualitative summaries based on the respective papers ($+$: favorable; $\triangle$: mixed; $-$: unfavorable).} %
\label{tab:positioning}
\setlength{\tabcolsep}{2pt}
\begin{tabular*}{\textwidth}{@{\extracolsep{\fill}}lllc>{\columncolor{AccessPatternShade}}l>{\columncolor{AccessPatternShade}}cl@{}}
\toprule
\textbf{Method} & \textbf{System model} & \textbf{Quality} & \textbf{Cost} & \textbf{Access pattern protection} & \textbf{Query protection} & \textbf{Database protection} \\
\midrule
Plaintext ANN (\eg SuCo~\cite{wei2025suco}) & Single-server & $+$ & $+$ & None & None & None \\
Path ORAM~\cite{DBLP:conf/ccs/StefanovDSFRYD13} & Single-server & --- & $-$ & Block addresses & --- & From SP \\
PP-ANNS~\cite{liu2025ppanns} & Single-server & $+$ & $\triangle$ & No trace guarantee & Yes & From SP \\
PACMANN~\cite{DBLP:conf/iclr/ZhouSF25} & Single-server & $\triangle$ & $\triangle$ & Private graph reads & Yes & Not required (public database) \\
BiSON~\cite{DBLP:conf/eurosp/DasRCG26} & Single-server & $\triangle$ & $\triangle$ & Oblivious graph reads & Yes & From SP \\
Panther~\cite{DBLP:conf/ccs/LiHZ0L0C25} & Single-server & $+$ & $\triangle$ & Private selection and reads & Yes & From user \\
FedKNN~\cite{DBLP:journals/pacmmod/ZhangWXPX24} & Multi (federated) & $+$ & $\triangle$ & SP-side protection & None & Across SPs \\
Preco~\cite{DBLP:conf/sp/Servan-Schreiber22} & Multi (two) & $+$ & $\triangle$ & Private bucket reads & Yes & Bounded leakage to user \\
\textbf{$\name$ (ours)} & Single-server & $+$ & $\triangle$ & obliviousness & Yes & Not required (public database) \\
\bottomrule
\end{tabular*}
\end{table*}

Table~\ref{tab:positioning} compares representative nearest neighbor search methods in terms of their system models, privacy targets, and performance. PP-ANNS~\cite{liu2025ppanns} protects query and database contents using encrypted comparisons, but does not establish a formal privacy guarantee for query-dependent access patterns. PACMANN~\cite{DBLP:conf/iclr/ZhouSF25} protects queries over a public database by letting the user traverse a graph and privately retrieve information at each hop. Panther~\cite{DBLP:conf/ccs/LiHZ0L0C25} protects the user's query from the server and the server's database from the user. BiSON~\cite{DBLP:conf/eurosp/DasRCG26} protects both the query and the outsourced database from the storage server and uses disk-compatible oblivious RAM (ORAM) to hide graph accesses. In the federated setting, FedKNN~\cite{DBLP:journals/pacmmod/ZhangWXPX24} protects the participating providers' data during collaborative $k$NN search, but its threat model does not require the user's query to be hidden from those providers.

Like PACMANN, we study asymmetric ANN search in the single-server setting. This raises a natural question: \emph{can we answer such queries more efficiently while providing a formal guarantee of query privacy against access-pattern leakage?}

\stab
\noindent\textbf{Our Approach.} To answer this question, our idea is to design an \emph{oblivious}\footnote{The notions of obliviousness are formally defined in Section~\ref{subsec:system}.} search framework that hides query-dependent memory accesses from SP. This goal leads to two challenges.

\noindent\emph{\textbf{Challenge 1:} An efficient oblivious search framework.} Efficient ANN search relies on indexes to restrict the search to promising candidates. However, incorporating an index into an oblivious framework requires protecting the entire search process. For example, graph traversal and cluster probing use query-dependent addresses, while candidate aggregation and refinement can reveal which items were selected. Protecting these operations without losing the efficiency of indexed search is challenging. Recent advances in subspace-collision search, including SuCo~\cite{wei2025suco} and TaCo~\cite{DBLP:journals/pacmmod/WeiLHXYP26}, offer an opportunity to design an efficient oblivious search framework. Building on their separation of cell probing, collision-score aggregation, and refinement, we jointly design $\name$, a privacy-preserving asymmetric ANN search framework. $\name$'s search base consists of a K-means-based inverted multi-index and fully oblivious ($\FO$) search~procedures: the index supports uniform cell retrieval, candidate activation~selects promising items within a public probe budget, and refinement uses their collision scores to identify the final results, while hiding query-dependent accesses throughout the search. These stages execute within a trusted execution environment, which protects private data values during computation, thereby combining data confidentiality with access-pattern protection to preserve query privacy throughout the search.

\noindent\emph{\textbf{Challenge 2:} Exploiting parallelism while preserving obliviousness.} Parallel execution is a natural direction for further reducing the query latency of $\name$'s $\FO$ search base.~However, parallelizing oblivious sorting alone does not address the dependencies across search stages: candidate activation is independent across subspaces, whereas collision-score accumulation updates shared state and determines subsequent candidate selection. Directly executing these updates concurrently requires coordination of both the scores and the ORAM state, which can limit parallelism. We therefore separate query-dependent local computation from global aggregation in public item order. Each subspace independently activates candidates and accumulates local collision scores, which are then combined through an oblivious merge to support global selection and refinement. Together with parallel oblivious sorting and distance computation, this design exposes parallelism while keeping worker assignments, access schedules, and synchronization points independent of the query to achieve the guarantee of full obliviousness. It preserves the search base's scoring semantics and retrieval quality at the cost of additional storage and aggregation work.

\stab
\noindent\textbf{Contributions.} Our contributions are as follows:
\begin{itemize}[leftmargin=3ex]
\item We formulate single-server asymmetric ANN search over a public vector database with a private query, and specify full obliviousness under this setting.
\item We propose $\name$, a privacy-preserving asymmetric ANN search framework with a subspace-collision-based search base, whose K-means-based inverted multi-index, candidate activation, and candidate refinement are jointly designed to protect query-dependent access patterns throughout the search. We prove that $\name$'s search base satisfies the privacy guarantee of full obliviousness.
\item We develop parallel $\name$ by decomposing the shared collision-score updates during the candidate refinement into independent local accumulation and oblivious global aggregation, enabling concurrent processing across subspaces without contention on a shared score ORAM. Together with parallel candidate activation, oblivious selection, and distance computation, this design exploits parallelism across the online search stages. We prove that it preserves full obliviousness and the search base's results, and analyze the trade-offs among total work, parallel running time, and auxiliary space.
\item We implement a prototype of $\name$ and evaluate its retrieval quality, search time, and index construction on six standard ANN benchmarks, with comparisons to non-private and privacy-preserving baselines. The experimental results demonstrate $\name$'s competitive performance.
\end{itemize}

\noindent\textbf{Organization}.
The rest of this paper is organized as follows. We introduce the background in Section~\ref{sec:background}. Section~\ref{sec:method} presents the overall framework of $\name$ together with $\name$'s $\FO$ search base. Then, we present {$\name$'s parallel optimization strategies in Section~\ref{sec:parallel}}.  
The experimental evaluation is presented in Section~\ref{sec:experiment} and the related work is discussed in Section~\ref{sec:related}.
We conclude this paper in Section~\ref{sec:conclusion}. 

\section{Background}\label{sec:background}
We first introduce the preliminaries in Section~\ref{subsec:preliminaries}. We then show the system model and security basis in Section~\ref{subsec:system}, and summarize SuCo's insecure subspace-collision pipeline for high-dimensional ANN search in Section~\ref{subsec:suco}. Some frequently used notations are summarized in Table~\ref{tab:notation}.

\begin{table}[t]
\centering
\small
\caption{Frequently used notations.}
\label{tab:notation}
\setlength{\tabcolsep}{2pt}
  \resizebox{1.0\linewidth}{!}{
\begin{tabular}{@{}p{0.28\columnwidth}p{0.8\columnwidth}@{}}
\toprule
\textbf{Notation} & \textbf{Meaning} \\
\midrule
$X=(x_1,\ldots,x_d)$ & a $d$-dimensional database vector \\
$\mathcal{X}\in\mathbb{R}^{n\times d}$ & vector database $\{X_1,\ldots,X_n\}$ \\
$q$ & private query vector in $\mathbb{R}^{d}$ \\
$\rho(\cdot,\cdot)$ & distance function for vector similarity search \\
$k,E_k(q),A_k(q)$ & query parameter, exact top-$k$ answer set, and approximate top-$k$ answer set \\
$\Pi$, $\mathcal{A}$, $\mathcal{L}$ & protocol, polynomial-time adversary, and common public information \\
$\access^\Pi(q, \mathcal{X})$ & the access pattern of operations that defines $\mathcal{A}$'s observation during the execution of $\Pi$ on $\mathcal{X}$ for $q$ \\
$\hat{\alpha},\beta$ & collision parameter and re-rank ratio \\
{$\alpha$} & collision ratio for SuCo \\
{$\mathsf{SC},\mathsf{SC}_i$} & {global subspace-collision score array and local score array for subspace $S_i$, respectively, where entry $j$ stores the score of item $X_j$} \\
{$R_i$, $\mathcal{R}$} & {activated identifier multiset for $S_i$, collection of these multisets}\\
$\mathcal{S},N_s$ & public subspace schedule and number of subspaces \\
$S_i,P_i,S_i^{(r)},P_i^{(r)}$ & subspace, projection function, and half-subspace/half-projection used by IMI \\
{$\sqrt{\kappa},\kappa,t$} & {number of K-means clusters, IMI cell count per subspace, and K-means iteration count} \\
{$C_i^{(r)},a_i^{(r)},q_i^{(r)}$} & {centroid set, assignment array, and projected query in half $r\in\{1,2\}$ of $S_i$}\\
$a_{i}^{(r)}[j]$ & assignment of the $j^\textnormal{th}$ item  $X_j$ in $\mathcal{X}$ {\wrt} the $r^\textnormal{th}$ ($r=1,2$) half dimensions in subspace $S_i$ \\
$Max[i], \bot$  & the maximum cell size of the inverted multi-index $IMI_i$ for subspace $S_i$, a unique virtual item identifier \\
{$IMI_i,IMIs$} & {inverted multi-index for $S_i$, list of these indexes} \\
{$p, \ell_i,\ell$ } & {number of workers per subspace, padded activation size $\ell_i=\hat\alpha\cdot Max[i]$, and total activation size $\ell=\sum^{N_s}_i\ell_i$} \\
\bottomrule
\end{tabular}
}
\end{table}

\subsection{Preliminaries}\label{subsec:preliminaries}
\noindent\textbf{Vector Similarity Search.}
Let $\mathcal{X}=\{X_1,\ldots,X_n\}$ be a database of $n$ vectors, where each vector item $X=(x_1,\ldots,x_d)\in\mathbb{R}^{d}$ has $d$ coordinates. We also use $\mathcal{X}\in\mathbb{R}^{n\times d}$ for its matrix representation, with row $j$ corresponding to item $X_j$. Given a query vector $q\in\mathbb{R}^{d}$ in the same metric space, similarity search retrieves database items according to a distance function $\rho$. 
In this paper, we consider the Euclidean distance function
\begin{equation}
\nonumber
    \rho(q,X)=\lVert q-X\rVert_2
    =\left(\sum_{\ell=1}^{d}(q[\ell]-x_\ell)^2\right)^{1/2},
    \label{eq:ts-euclidean}
\end{equation}
where $q[\ell]$ is the $\ell$-th query coordinate. Then, given $1\leq k\leq n$, \emph{exact} top-$k$ search returns a set $E_k(q)$ of $k$ vectors in $\mathcal{X}$ nearest to $q$. That is,

\begin{equation}
\nonumber
    E_k(q)
    =
    \arg\min_{\substack{\mathcal{X}'\subseteq \mathcal{X},~|\mathcal{X}'|=k}}
    \sum_{X_i\in \mathcal{X}'}\rho(q, X_i).
    \label{eq:exact-ts-topk}
\end{equation}

\noindent\textbf{Approximate Top-$k$ Nearest Neighbor ($\ann$) Search.}
To enable efficient retrieval, \emph{approximate top-$k$ nearest-neighbor} search returns a set $A_k(q)$ of $k$ database vectors, aiming for high overlap with $E_k(q)$ while avoiding an exhaustive search for the exact answers. Following existing $\ann$ works~\cite{wei2025suco, DBLP:conf/iclr/ZhouSF25, liu2025ppanns}, we adopt the metric of recall to evaluate the utility of an approximate search result $A_k(q)$. That is,
\begin{equation}
    \begin{aligned}
    \operatorname{Recall@}k(q)
    &=
    \frac{\big|A_k(q)\cap E_k(q)\big|}{k}.
    \end{aligned}
    \label{eq:ann-recall}
\end{equation}

Then, we introduce the notions of subspace and projection used for subspace collision-based $\ann$~\cite{wei2025suco, DBLP:journals/pacmmod/WeiLHXYP26}.

\stab
\noindent\textbf{Subspaces and Projections.}
Let $\mathcal{S}=\{S_1,\ldots,S_{N_s}\}$ partition $d$ coordinate dimensions into $N_s$ nonempty, disjoint groups. The projection $P_i$ selects the coordinates in $S_i$, giving
\begin{equation}
\nonumber
    q_i=P_i(q),\qquad
    \mathcal{X}_i=\{P_i(X_j):1\leq j\leq n\}.
    \label{eq:subspace-projection}
\end{equation}

For the $N_s$ disjoint subspaces, squared Euclidean distance satisfies
\begin{equation}
\nonumber
    \lVert q-X_j\rVert_2^2
    =\sum_{i=1}^{N_s}\lVert q_i-P_i(X_j)\rVert_2^2.
    \label{eq:distance-decomposition}
\end{equation}

Projected distance rankings may differ across subspaces. The subspace-collision approach in Section~\ref{subsec:suco} therefore efficiently aggregates rank evidence  from multiple projections via distance-based indexes in subspaces  to select candidates, and uses the original-space distance for final ranking.

\subsection{System Model and Security Basis}
\label{subsec:system}
\noindent\textbf{System Model.}
We consider a cloud computing scenario with query privacy protection: the user submits a private query vector to a service provider (SP) for $\ann$ retrieval over a public vector database and its query-independent index.
We adopt the widely used assumption in cloud computing that SP is an \emph{honest-but-curious} adversary: it follows the computational protocol but may infer sensitive information about the user's private query. Under this model, we~consider the following two kinds of privacy as the \emph{privacy target}:
\begin{itemize}[leftmargin=3ex]
\item \textbf{Data privacy.} The data privacy is to protect the user's query vector and any query-dependent intermediate values except for the query parameter $k$; and
\item \textbf{Access pattern privacy.} The access pattern privacy is to ensure that $\SP$ cannot infer any information that may compromise the data privacy. 
\end{itemize}

Next, we introduce the security basis used in this paper.

\stab
\noindent\textbf{Security Basis.}
To protect data privacy, there have been cryptographic techniques supporting secure computational operators, such as secret sharing~\cite{DBLP:journals/cacm/Shamir79, DBLP:conf/codcry/Beimel11}, homomorphic encryption scheme~\cite{DBLP:conf/stoc/Gentry09, DBLP:journals/toct/BrakerskiGV14, DBLP:conf/asiacrypt/CheonKKS17}, and trusted execution environment (TEE)~\cite{IntelTDX2022, AMDMemoryEncryption2021, NVIDIAConfidentialComputing2023}. To achieve the above privacy target, we can leverage these techniques to preserve the data privacy and mainly focus on designing secure search protocols with protection on the \emph{access pattern privacy}. 

To protect access pattern privacy, obliviousness is a standard security notion that provides formal privacy guarantee. 
{We adopt full obliviousness~\cite{DBLP:conf/stoc/Goldreich87} to protect the complete observable search execution pattern.}

\begin{definition}[\textbf{Full Obliviousness}]
\label{def:fo}
Given a public vector database $\mathcal{X}$, its index, and the public algorithm parameters, including $d$ and $k$, denote this common public information by $\mathcal{L}$. Then, for any two query vectors $q,q'\in\mathbb{R}^{d}$ with the same public parameters, a protocol $\Pi$ is \emph{fully oblivious (or simply $\FO$)} if, for every probabilistic polynomial-time adversary $\mathcal{A}$, there exists a negligible function $\delta(\lambda)$ such that
\begin{align}
\label{eqa:fo}
&\Pr\Big[\mathcal{A}\big(\mathcal{L},\access^\Pi(q,\mathcal{X})\big)=1\Big] \nonumber\\
&\quad\leq\Pr\Big[\mathcal{A}\big(\mathcal{L},\access^\Pi(q',\mathcal{X})\big)=1\Big]+\delta(\lambda).
\end{align}
where (i) $\access^\Pi(q,\mathcal{X})$ denotes the observable access trace of executing $\Pi$ on $q$ and $\mathcal{X}$; (ii) $\mathcal{A}$ outputs a bit to distinguish the two trace distributions. The probabilities are over the randomness of $\Pi$ and $\mathcal{A}$; and (iii) $\lambda$ is the cryptographic security parameter, and $\delta(\lambda)$ is a negligible upper bound on the adversary's distinguishing advantage between the observable access traces induced by $q$ and $q'$ under the same public parameters.
\end{definition}

Now, we are ready to present the problem statement.

\stab
\noindent\textbf{Problem Statement.}
Given the database $\mathcal{X}$ and a private user query $q$ with parameter $k$, the goal is to design secure $\ann$ search approaches with access pattern privacy preservation under the system model above, while the data privacy can be protected via standard cryptographic techniques, \eg TEE.

\subsection{Pipeline for Subspace Collision-Based $\ann$ Search}
\label{subsec:suco}

\begin{figure}[!t]
\centering
\includegraphics[width=\columnwidth]{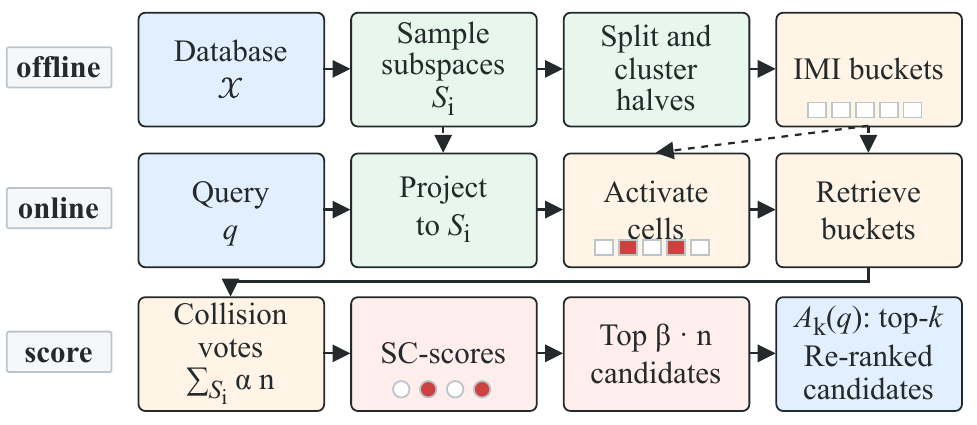}
\caption{Plaintext SuCo pipeline from IMI construction to top-$k$ reranking.}
\Description{A three-row pipeline for plaintext SuCo. The offline row builds sampled subspaces, clustered halves, and inverted multi-index buckets from the database. The online row projects the query, activates cells, and retrieves buckets. The scoring row gathers collision votes, aggregates SC-scores, keeps top candidates, and reranks them to return the approximate top-k answer.}
\label{fig:plain-suco-overview}
\end{figure}

Figure~\ref{fig:plain-suco-overview} summarizes SuCo's plaintext pipeline~\cite{wei2025suco} that consists of offline index construction followed by online search.

\begin{itemize}[leftmargin=*,labelindent=0pt,labelsep=0.5em,align=left]
\item \textbf{Offline index construction.} SuCo partitions the $d$ dimensions into $N_s$ sampled subspaces and builds an inverted multi-index (IMI) for each subspace $S_i$. It then splits $S_i$ into two halves and clusters each half into $\sqrt{\kappa}$ centroids using K-means. Each identifier is stored in one of $\kappa$ IMI cells according to its nearest centroid pair.

\item \textbf{Online search.} SuCo projects  the user query $q$ into each subspace and computes its distances to both centroid sets. Then, it prioritizes IMI cells by the sums of these distances and retrieves identifiers until at least $\alpha\cdot n$ items are activated as candidates, where $\alpha$ is the collision ratio and $n$ is the database size. Next, SuCo (i) counts one collision vote for each retrieved item per subspace, and (ii) retains the $\beta\cdot n$ highest-scoring candidates, where $\beta$ is the re-rank ratio. Finally, SuCo re-ranks these $\beta\cdot n$ candidates by original-space distances to return the closest $k$ ones as $A_k(q)$.
\end{itemize}

SuCo's index cell retrieval, candidate activation, and refinement accesses depend on the user query $q$ and expose its access pattern. In Section~\ref{sec:method}, we present our $\name$ search framework that protects $q$ against access-pattern leakage. 

\begin{figure*}[t]
\centering
\includegraphics[width=1\textwidth]{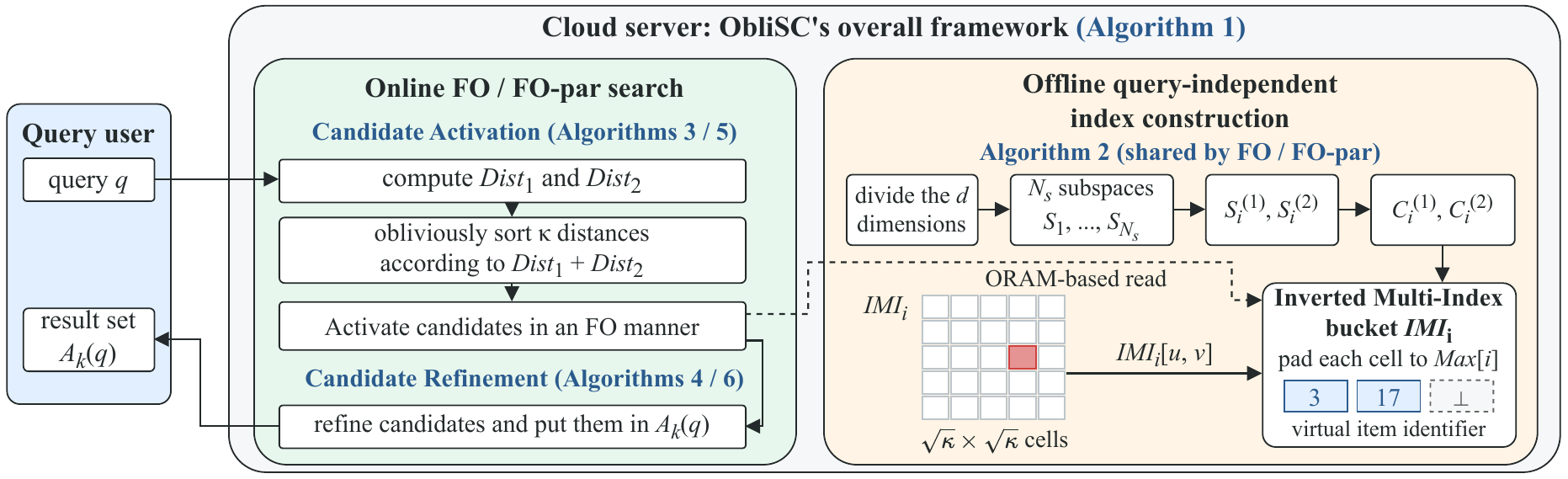}
\caption{$\name$ system model: offline query-independent index construction and online $\FO$ search processing.}
\Description{A query user sends a private query to the cloud server and receives the result set. Sequential FO and parallel FO search share the offline index constructed by Algorithm 2. Online activation uses Algorithm 3 or 5, and refinement uses Algorithm 4 or 6. The offline panel divides dimensions into subspaces, constructs centroid groups and IMI buckets, and illustrates a padded cell containing real and virtual item identifiers.}
\label{fig:ppsc-overview}
\end{figure*}

\section{$\name$'s Overall Search Framework}
\label{sec:method}

This section illustrates the overview and the fully oblivious search base of our subspace collision-based $\name$ search framework. Due to space limitations,~some detailed proofs are presented in Appendix~A of the technical report~\cite{oblisc:tr}.

\subsection{$\name$'s Overview}
\label{subsec:ppsc-overview}

$\name$ has three components: (i) one-time offline \emph{query-independent} inverted multi-index (IMI) construction, (ii) fully oblivious ($\FO$) IMI based candidate activation, and (iii) $\FO$ top-$k$ refinement. 
Figure~\ref{fig:ppsc-overview} shows the query user and the cloud server that runs $\name$'s overall framework, while Algorithm~\ref{alg:ppsc} gives the corresponding procedure.

\begin{figure}[!t]
\centering
\includegraphics[width=\columnwidth]{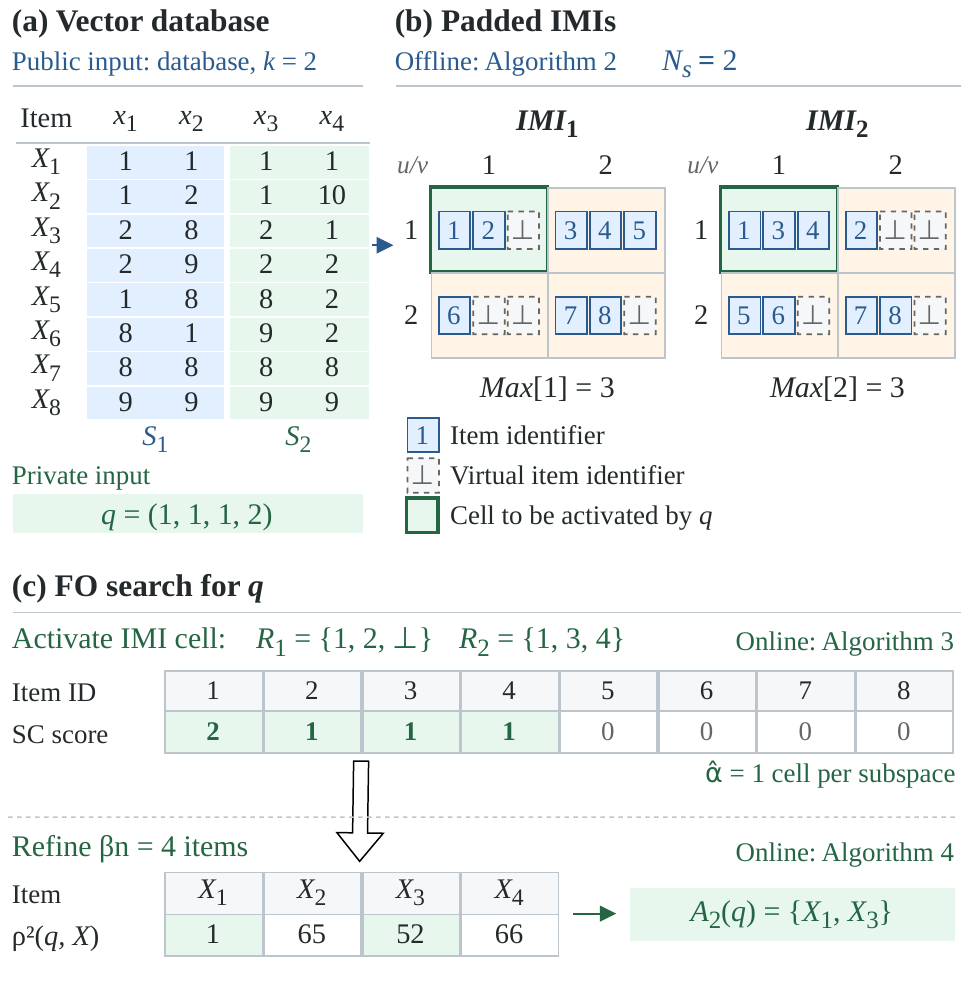}
\caption{A running example of $\name$.
}
\label{fig:ppsc-running-example}
\Description{Eight four-dimensional vectors are partitioned into two subspaces. Each two-by-two IMI has three identifier entries per cell after padding. Query q activates cell (1,1) in both IMIs, producing collision scores (2,1,1,1,0,0,0,0). Four candidates are refined and X1 and X3 are returned. A different query selects different cells under the same public operation counts.}
\end{figure}

\begin{algorithm}[tbp]
\small
\DontPrintSemicolon
\SetKwFunction{comOne}{IndexConstruct}
\SetKwFunction{comTwo}{CanActive}
\SetKwFunction{comThree}{Top$k$Refine}
\SetKwFunction{comFour}{{pending}}
\KwIn{database $\mathcal{X}$ with $n$ $d$-dimensional items, query $q$ with parameter $k$, subspace number $N_s$, collision parameter $\hat{\alpha}$, and re-rank ratio $\beta$}
\KwOut{approximate top-$k$ answer set $A_k(q)$}
\BlankLine

\tcp*[h]{{(i) Offline Query-Independent Index Construction}}

\comOne$(\cdot)$ 
\hfill \tcp*[h]{generate IMI index}

\BlankLine
\tcp*[h]{Online Search Process}

initialize $\mathsf{SC}[j]\leftarrow0$ for each item $X_j$, $j\in[1, n]$ in $\mathcal{X}$\;

divide the $d$ dimensions into $N_s$ subspaces $S_1,\ldots,S_{N_s}$
and project $q$ into every subspace\;

initialize multiset $\mathcal{R}\gets \emptyset$ of elements in activated cluster sets containing answers\;

\For{$i \leftarrow 1$ \KwTo $N_s$}{

    \tcp*[h]{(ii) $\FO$ Candidate Activation}%

    $\mathcal{R}\gets \mathcal{R}~\cup$ $\big\{$\comTwo$(\cdot)\big\}$\;

}

\BlankLine

    \tcp*[h]{(iii) $\FO$ Top-$k$ Refinement: count the $\mathsf{SC}$ and refine the top-$k$ answers}

    $A_k(q)\gets$ \comThree$(\cdot)$ 

\Return $A_k(q)$\;
\caption{$\name$'s overall framework}
\label{alg:ppsc}
\end{algorithm}

Taking as inputs the database $\mathcal{X}$ with $n$ $d$-dimensional vector items, a query $q$ with parameter $k$, the subspace number $N_s$, the collision parameter $\hat{\alpha}$, and the re-rank ratio $\beta$,  Algorithm~\ref{alg:ppsc} outputs an approximate top-$k$ answer set $A_k(q)$. First, as $\name$'s component~(i),  Line~1 invokes the {\tt\small IndexConstruct} procedure to generate the one-time offline subspace collision-based IMI index. Then, Lines~2-4 initialize the online query process for the input query $q$ by setting the \emph{subspace collision} (SC) score $\mathsf{SC}$ of each data item as $0$, and dividing and projecting $q$ into each of the $N_s$ subspaces via projections. For each subspace, Lines~5-6 invoke the {\tt\small CanActive} procedure as $\name$'s component~(ii)  to obtain the activated candidate item (identifier) set by using IMI  in an $\FO$ manner to find promising items. The elements in these sets are inserted into multiset $\mathcal{R}$. Line~7 invokes the $\FO$ {\tt\small Top$k$Refine} procedure as $\name$'s component~(iii) to refine the top-ranked candidates in $\mathcal{R}$ and obtain an $\ann$ answer set $A_k(q)$, which is returned as the search result in Line~8.
The following subsections will illustrate each of these three components.

\begin{algorithm}[tbp]
\small
\DontPrintSemicolon
\KwIn{database $\mathcal{X}$ with $n$ $d$-dimensional items, subspace number $N_s$, {{IMI cell count per subspace} $\kappa$}, and K-means iterations $t$}
\KwOut{centroid list $centroids$, IMI list $IMIs$, and maximum cluster size array $Max$}
\BlankLine

\tcp*[h]{{Initialization}}

initialize $centroids$ with length $2N_s$, $IMIs$ with length $N_s$\;
divide the $d$ dimensions into $N_s$ subspaces $S_1,\ldots,S_{N_s}$\;
divide all items into subspace datasets $\mathcal{X}_1,\ldots,\mathcal{X}_{N_s}$\;

\BlankLine

\tcp*[h]{{\textsc{KMeans}-Based Index Construction}}

\For{$i \leftarrow 1$ \KwTo $N_s$}{
    initialize a map $IMI_i$ as the inverted multi-index\;
    split $S_i$ into $S_i^{(1)}$ and $S_i^{(2)}$,  split $\mathcal{X}_i$ into $\mathcal{X}_i^{(1)}$ and $\mathcal{X}_i^{(2)}$\;
    $(C_i^{(1)},a_i^{(1)})\leftarrow \textsc{KMeans}(\mathcal{X}_i^{(1)},\sqrt{\kappa},t)$\;
    $(C_i^{(2)},a_i^{(2)})\leftarrow \textsc{KMeans}(\mathcal{X}_i^{(2)},\sqrt{\kappa},t)$\;
    append $C_i^{(1)}$ and $C_i^{(2)}$ to $centroids$\;

    \For{$j \leftarrow 1$ \KwTo $n$}{
        $u\leftarrow a_i^{(1)}[j]$, $v\leftarrow a_i^{(2)}[j]$\;
        $IMI_i[u,v].append(j)$
\hfill \tcp*[h]{$IMI_i$: $\sqrt{\kappa}\times \sqrt{\kappa}$ cells}
    }
    {$Max[i] \gets$ the maximum cell size of $IMI_i$\;}
    {pad all cells in $IMI_i$ to size $Max[i]$ using a unique virtual item identifier $\bot$\;}
    append $IMI_i$ to $IMIs$\;
}
\Return $\big\langle centroids, IMIs, {Max}\big\rangle$\;
\caption{The {\tt\small IndexConstruct$(\cdot)$} procedure}
\label{alg:index}
\end{algorithm}

\subsection{SC-based Index Construction}
\label{subsec:indexconstruction}
According to the existing SC-based $\ann$ search methods~\cite{wei2025suco, DBLP:journals/corr/abs-2603-24919}, the IMI construction depends on the database $\mathcal{X}$ only, which is independent of the input query. However, these methods do not consider the fixed-shape index design, which is critical for achieving $\FO$ online search since query-dependent online search processing may access index entries of different sizes, compromising access-pattern privacy. 
Therefore, we propose to construct fixed-shape IMI to facilitate the oblivious design for $\name$'s online search. In Figure~\ref{fig:ppsc-overview}, the rectangle on the right illustrates $\name$'s IMI layout for a single subspace and the padded cell representation.

The IMI construction is shown in Algorithm~\ref{alg:index}. Taking as inputs the database $\mathcal{X}$ with $n$ $d$-dimensional items, the subspace number $N_s$, the number of K-means clusters $\sqrt{\kappa}$ (or the IMI cell count per subspace $\kappa$), and the K-means iterations $t$, the {\tt\small IndexConstruct} procedure outputs as IMI the centroid list $centroids$, the IMI list $IMI$s, and an array $Max$ of each subspace's maximum cluster size used for $\FO$ design.
First, Line~1 initializes \emph{centroids} with length $2\cdot N_s$ and \emph{IMI}s with length $N_s$, and Line~2 divides the $d$ dimensions into $N_s$ subspaces
via \emph{uniformly sampling without replacement}. Accordingly, Line~3 divides all items into $N_s$ subspace datasets $\mathcal{X}_1,\ldots,\mathcal{X}_{N_s}$.
Then, Lines~4-14 construct the IMI $IMI_i$ for the $i^\textnormal{th}$ ($1\le i\le N_s$) subspace $S_i$ as follows: (i) Line~5 initializes a 2-dimensional map as the initial $IMI_i$; (ii) Line~6 splits $S_i$ into two disjoint and lower-dimensional halves $S_i^{(1)}$ and $S_i^{(2)}$ \wrt the first $\lfloor d_i/2\rfloor$ and the remaining $d_i-\lfloor d_i/2\rfloor$ dimensions, where $d_i=\dim(S_i)=d/N_s$, respectively. Similarly, each subspace dataset $\mathcal{X}_i$ is also split into two halves $\mathcal{X}_i^{(1)}$ and $\mathcal{X}_i^{(2)}$; (iii) for the first (\resp the second) half dimensions of $S_i$, Line~7 (\resp Line~8) invokes \emph{K-means} on $\mathcal{X}_i^{(1)}$ (\resp $\mathcal{X}_i^{(2)}$) to compute  the centroid set $C_i^{(1)}$ (\resp $C_i^{(2)}$) of $\sqrt{\kappa}$ centroids in $S_i$ together with the assignment {array} $a_i^{(1)}$ (\resp $a_i^{(2)}$) of $n$ assignments (\ie the \emph{nearest} centroid to each of the $n$ items). Line~9 appends $C_i^{(1)}$ and $C_i^{(2)}$ in \emph{centroids}; (iv) Lines~10-12 append identifier $j$ to the 2-dimensional $IMI_i$ with index key pair $\big(a^{(1)}_i[j], a^{(2)}_i[j]\big)$ for the $j^\textnormal{th}$ ($1\le j\le n$) item; and (v) for the purpose of $\FO$ design, Line~13 stores in $Max[i]$ the maximum cell size of $IMI_i$, and then Line~14 pads each cell in $IMI_i$ to size $Max[i]$ using a unique virtual item identifier $\bot$, which indicates a virtual item having $+\infty$ distances from any queries.
Finally, Line~16 returns $\big\langle centroids, IMIs, Max\big\rangle$.

\begin{example}[IMI Index Construction]
\label{example:index-construction}
Consider a running example of a database that consists of the eight four-dimensional vectors in Figure~\ref{fig:ppsc-running-example}(a), with a private query $q=(1,1,1,2)$ and $k=2$. Let $N_s=2$ with coordinate groups $S_1=\{1,2\}$ and $S_2=\{3,4\}$, let $\kappa=4$, giving two centroids per half-subspace, and let the K-means iterations $t$ $=$ $2$.

For each one-dimensional half-subspace, the IMI construction invokes K-means for $2$ iterations. The resulting assignments give the IMIs in Figure~\ref{fig:ppsc-running-example}(b). For instance, $IMI_1[1,1]$ contains identifiers $1,2$, whereas $IMI_1[1,2]$ contains $3,4,5$. Both IMIs have maximum cell size three, so $Max[1]=Max[2]=3$. Padding changes $IMI_1[1,1]$'s cell to $(1,2,\bot)$.
\end{example}

\noindent\textbf{Analysis.}
$centroids$, $IMIs$, and $Max$ provide efficient search structure used by the later candidate activation procedure, and their constructions  depend on database $\mathcal{X}$ and public parameters only. That is, the input query $q$ has no effect on the access pattern $\access^\Pi(q, \mathcal{X})$ of {\tt\small IndexConstruct}. According to Definition~\ref{def:fo}, we can easily know that {\tt\small IndexConstruct} satisfies full obliviousness.

\subsection{$\FO$ Candidate Activation}
The existing activation procedure dynamically activates the candidate clusters that contain the query answers, which involves a lot of query-dependent code branches and hence fails to satisfy full obliviousness. To handle this, we propose an $\FO$ candidate activation procedure.

\begin{algorithm}[tbp]
\small
\DontPrintSemicolon
\KwIn{Subspace $S_i$, projected query $q_i$,  collision parameter $\hat{\alpha}$, dataset size $n$, {IMI cell count per subspace} $\kappa$, and $\big\langle centroids, IMIs,Max\big\rangle$}
\KwOut{Multiset $R$ of identifiers for activated items whose item points collide with $q_i$ in $S_i$}
\BlankLine

\tcp*[h]{Initialization for $q_i$}

compute array $Dist_{1}$ (\resp $Dist_{2}$) of distances between $q_i^{(1)}$ (\resp $q^{(2)}_i$)  and each of the $\sqrt{\kappa}$ centroids in subspace $S^{(1)}_i$ (\resp $S^{(2)}_i$)\;

{invoke {\tt\small Obli\_Sort} to sort all $\kappa$ pairwise distance sums $Dist_1[u]+Dist_2[v]$ in ascending order and store the corresponding centroid identifier pairs $(u,v)$ in array $ID$ in the same order}\;

\BlankLine

\tcp*[h]{Candidate Activation}

$R\gets \emptyset$\;

\For{$pos$ $\leftarrow 1$ \KwTo $\hat{\alpha}$}{

$u, v \gets$ $ID[pos]$\;

$cluster \gets IMI_i[u, v]$  \hfill \tcp*[h]{ORAM-based read}

$R\gets R\uplus cluster$ \hfill \tcp*[h]{$\uplus$: multiset union}
}

\Return $R$\;
\caption{$\FO$ {\tt\small CanActive$(\cdot)$} procedure}
\label{alg:canactive}
\end{algorithm}

As shown in Algorithm~\ref{alg:canactive}, taking as inputs the subspace $S_i$, the projected query $q_i$ in $S_i$,  the collision parameter $\hat{\alpha}$, the dataset size $n$, the IMI budget $\kappa$, and $\name$'s index structure $\big\langle centroids$, $IMIs$, 
$Max\big\rangle$, the {\tt\small CanActive} procedure outputs the activated candidate (item identifier) multiset $R$ for $S_i$. 
First, Line~1 computes and stores in arrays $Dist_1$ and $Dist_2$ the distances from the projected queries $q_i^{(1)}$ and $q_i^{(2)}$ (\ie the projected queries of $q_i$ \wrt the first and second half dimensions in $S_i$)  to the two centroid groups $C_i^{(1)}$ and $C_i^{(2)}$, respectively.  Then, Line~2 invokes the data-oblivious sorting procedure {\tt\small Obli\_Sort} to sort all $\kappa$ pairwise distance sums in ascending order and stores the corresponding centroid identifier pairs in array $ID$ in the same order.
To activate candidate clusters, the existing subspace-collision pipeline  uses a collision ratio $\alpha$ and dynamically retrieves at least $\alpha\cdot n$ items from the activated clusters. Different from such a query-dependent dynamic activation, {\tt\small CanActive} adopts an input collision parameter $\hat{\alpha}$ as a fixed number of $IMI_i$ cell accesses. This ensures that Lines~4-7 always retrieve
$Max[i]$ identifiers from each of the $\hat{\alpha}$ cells corresponding to the first $\hat{\alpha}$ centroid identifier pairs in $ID$. It is worth noting that Line~6 adopts a standard ORAM-based memory access approach $\big($\eg Path-ORAM~\cite{DBLP:conf/ccs/StefanovDSFRYD13} and  Circuit ORAM~\cite{DBLP:conf/ccs/WangCS15}$\big)$ to read the $Max[i]$ items stored in a specific cell of IMI bucket $IMI_i$, so as to protect the memory access pattern. Finally, Line~8 returns multiset $R$ of $\hat{\alpha}\cdot Max[i]$ activated item identifiers.

\begin{example}[$\FO$ Candidate Activation]
Following Example~\ref{example:index-construction}, we consider the candidate activation for query $q$ $=$ $(1,1,1,2)$. Assume that $\hat{\alpha}$ $=$ $1$. 
For $q$, the centroid pair with the smallest distance sum is $(1,1)$ in both IMIs. Therefore, as shown in Figures~\ref{fig:ppsc-running-example}(b)-\ref{fig:ppsc-running-example}(c), {\tt\small CanActive} reads $IMI_1[1, 1]$ and $IMI_2[1, 1]$ in an ORAM-based method to activate the candidates with identifiers in $R_1=\{1,2,\bot\}$ and $R_2=\{1,3,4\}$, respectively. 
\end{example}

\noindent\textbf{Analysis.} Throughout our complexity analysis, we treat ORAM as a black box and assume that accessing one of $N$ blocks, each containing $b$ scalar entries, takes $O(b\cdot\log N)$ time. Each padded IMI cell is one block of $Max[i]$ identifiers, each score is a constant-size block, and each database item is a block of $d$ coordinates.
Then, for the time complexity, {\tt\small CanActive} takes $O(\sqrt{\kappa}\cdot d/N_s)$ time to compute $Dist_1$ and $Dist_2$ (Line~1),\footnote{Assume that arithmetic operations take $O(1)$ time.} $O(\kappa \log^2 \kappa)$ time to obliviously sort $\kappa$ distances (Line~2), and $O\big(\hat{\alpha}\cdot Max[i]\cdot \log \kappa^2\big)$ 
time to obtain the activated candidate set $R$ (Lines~3-7). Hence, the total time complexity of invoking $N_s$ times of {\tt\small CanActive} is $O\big(\sqrt{\kappa}\cdot d + N_s\cdot\kappa \log^2 \kappa + \hat{\alpha}\cdot \log \kappa\cdot\sum^{N_s}_{i=1}{Max[i]}\big)$.

Our privacy analysis assumes protected private values and oblivious execution under fixed public inputs and parameters.\footnote{The TEE protects query values, intermediate values, and results, while query and result transfers remain confidential. We assume adaptively secure ORAM and oblivious local operations, including observable ORAM-internal accesses, with storage and per-access work padded to public bounds.}\label{para:privacy-assumptions}
An indistinguishability analysis of Algorithm~\ref{alg:canactive} under these assumptions yields the following lemma.

\begin{lemma}
\label{lemma:base}
{\tt\small CanActive} satisfies full obliviousness. 
\end{lemma}

\begin{proof}
Consider two inputs of {\tt\small CanActive} that have the same public index and parameters but different query data values. Comparing Algorithm~\ref{alg:canactive} for both inputs, we have the following: (i) Lines~1-2 compute a fixed number of distances and obliviously sort all $\kappa$ distance sums, so their access schedules are independent of the query values; and (ii) Lines~3-7 scan the first $\hat{\alpha}$ entries of $ID$ in the same order and retrieve $Max[i]$ identifiers per IMI cell. Although the selected cell addresses may differ, every retrieval uses ORAM and the retrieved identifiers are appended to $R$ in a fixed-length scan. Hence, both executions issue the same number of fixed-size ORAM requests. By adaptive ORAM security, the corresponding observable traces are computationally indistinguishable. The remaining access schedules are identical, and both outputs have size $\hat{\alpha}\cdot Max[i]$. Therefore, Equation~\eqref{eqa:fo} holds, and Lemma~\ref{lemma:base} follows.
\end{proof}

\subsection{$\FO$ top-$k$ Candidate Refinement}
After {\tt\small CanActive}, $\name$ aggregates the collision votes (\ie collided candidates) across subspaces and refines a fixed number of promising candidates in the original~space.

\begin{algorithm}[tbp]
\small
\DontPrintSemicolon
\KwIn{Subspace number $N_s$, multiset $\mathcal{R}$ of activated candidate item identifier in multisets $R_i$ for all subspaces $S_i$, dataset size $n$, query $q$ with parameter $k$, and re-rank ratio $\beta$}
\KwOut{Approximate top-$k$ answer set $A_k(q)$}

\BlankLine

\For{$i \leftarrow 1$ \KwTo $N_s$}{
    \ForEach{$point\_id\in R_i$}{
            $\mathsf{SC}[point\_id]++$  \hfill \tcp*[h]{ORAM-based update}
    }
}
invoke {\tt\small Obli\_Sort} to sort $\mathsf{SC}$ in descending order\;
$A\gets \emptyset$\;
\For{$z \leftarrow 1$ \KwTo $\beta\cdot n$}{
    let $\tau$ be the item identifier for the $z^\textnormal{th}$ element in $\mathsf{SC}$\;
    $X'\gets X_\tau$ \hfill \tcp*[h]{ORAM-based read}
    
    append $\langle\rho(X',q),\tau\rangle$ to the next public slot of $A$ \tcp*[h]{compute original distance}
}
invoke {\tt\small Obli\_Sort} to sort $A$ in ascending order of distance\;
retrieve the items of the first $k$ elements in sorted $A$ through ORAM and write them into consecutive slots of $A_k(q)$\;
\Return $A_k(q)$\;
\caption{$\FO$ {\tt\small Top$k$Refine} procedure}
\label{alg:refine}
\end{algorithm}

As shown in Algorithm~\ref{alg:refine}, taking as inputs the subspace number $N_s$, the multiset $\mathcal{R}$ of activated candidate (item) identifiers in multisets $R_1, ..., R_{N_s}$, the dataset size $n$, the query $q$ with parameter $k$, and the re-rank ratio $\beta$, the {\tt\small Top$k$Refine} procedure outputs the approximate top-$k$ answer set $A_k(q)$. First, Lines~1-3 aggregate subspace-collision votes by (i) linearly scanning the activated candidate identifiers in each $R_i$ ($1\le i \le N_s$) and (ii) incrementally increasing the $\mathsf{SC}$ of these identifiers through ORAM-protected read-modify-write operations to protect the memory access pattern. Note that each virtual identifier also triggers an ORAM update to a reserved dummy score slot, which is excluded from the subsequent sorting of real-item scores. Then, Line~4 invokes {\tt\small Obli\_Sort} to obliviously sort $\mathsf{SC}$ in descending order. The items having the first  $\beta\cdot n$ identifiers in the sorted $\mathsf{SC}$ are considered as the promising ones to be refined: for each of  these $\beta\cdot n$ candidates (Line~6), Line~7 obtains the selected identifier $\tau$, Line~8 retrieves the data of $X_\tau$ through ORAM-based reads, and Line~9 computes the original-space distance $\rho(X_\tau,q)$ and appends the distance--identifier pair to array $A$. Then, to refine these $\beta\cdot n$ candidates, Line~10 obliviously sorts the element pairs in $A$ by their distances in ascending order, and Line~11 retrieves the vectors in the first $k$ elements in sorted $A$  and inserts them into the answer set $A_k(q)$. Finally, Line~12 returns $A_k(q)$ as the output.

\begin{example}[$\FO$ Candidate Refinement]
Aggregating $R_1$ and $R_2$ gives real-item scores $(2,1,1,1,0,0,0,0)$. Virtual entries consume protected dummy operations without contributing to real-item scores, and have distance $+\infty$. The four highest-scoring items are $X_1,X_2,X_3,X_4$; their squared Euclidean distances to $q$ are $1,65,52,66$, respectively. Refinement therefore returns $A_2(q)=\{X_1,X_3\}$. This illustrates why collision scores select candidates but do not determine the final distance ranking. The four-item refinement count remains fixed for $q'$, while the selected identifiers and distances remain private.
\end{example}

\noindent\textbf{Analysis.} Let $\ell$ $=$ $\sum^{N_s}_{i=1}{\hat{\alpha}\cdot Max[i]}$.
{\tt\small Top$k$Refine} takes $O(\ell\cdot \log n )$ time to scan the activated items for oblivious write on $\mathsf{SC}$ (Lines~1-3), $O({n}\cdot\log^2 {n})$ time to obliviously sort $\mathsf{SC}$ (Line~4) and the elements in $A$ (Line~10), and $O\big(\beta\cdot n\cdot d\cdot\log n\big)$ time to obliviously read $\beta\cdot n$ data items and compute their corresponding original-space distances (Lines~6-9).
The vector data retrieval in Line~11 takes $O(k\cdot d\cdot\log n)$ time. Therefore, the total time complexity of {\tt\small Top$k$Refine} is $O\big(\ell\cdot \log n + {n}\cdot\log^2 {n} + \beta\cdot n\cdot d\cdot\log n\big)$.

An indistinguishability analysis of Algorithm~\ref{alg:refine} also establishes Lemma~\ref{lemma:two} for {\tt\small Top$k$Refine}'s access pattern.

\begin{lemma}
\label{lemma:two}
{\tt\small Top$k$Refine}  satisfies full obliviousness. 
\end{lemma}

Lemma~\ref{lemma:two} can be proved using an indistinguishability analysis similar to that for Lemma~\ref{lemma:base}. Consequently,
the analysis in Section~\ref{subsec:indexconstruction} and Lemmas~\ref{lemma:base}-\ref{lemma:two} establish full obliviousness for all three  procedures of $\name$, which covers {\tt\small IndexConstruct}, {\tt\small CanActive}, and {\tt\small Top$k$Refine}. As the access patterns of the operations in Lines~2-4 of Algorithm~\ref{alg:ppsc} are also independent of the query values, Theorem~\ref{theo:one} follows.

\begin{theorem}
\label{theo:one}
$\name$  satisfies full obliviousness. 
\end{theorem}

\section{Parallel Optimization}
\label{sec:parallel}
In this section, we further improve the execution of $\name$'s search base through parallelism on its candidate activation and refinement, while preserving full obliviousness.

\subsection{Parallel Candidate Activation}
\label{subsec:parallel-activation}
As the $d$ dimensions are divided into $N_s$ subspaces via uniform sampling \emph{without} replacement, each subspace contains a disjoint subset of dimensions. Given the index and projected query, candidate activation in each subspace does not depend on the results from other subspaces. Accordingly, we can directly parallelize candidate activation across subspaces.

\begin{algorithm}[tbp]
\small
\DontPrintSemicolon
\KwIn{Inputs of Algorithm~\ref{alg:canactive} for all $N_s$ subspaces, and public worker count $p$ per subspace}
\KwOut{Collection $\mathcal R=\{R_1,\ldots,R_{N_s}\}$ of activated identifier multisets}
{assign a fixed group of $p$ workers to each subspace $S_i$, initialize $R_i\gets\emptyset$}\;
\For{$i\gets1$ \KwTo $N_s$ \textnormal{in parallel}}{
{compute $Dist_1$ and $Dist_2$ in parallel using the $p$ workers assigned to $S_i$}\;

{invoke parallel {\tt\small Obli\_Sort} using these $p$ workers to sort the $\kappa$ distance sums in ascending order and store the corresponding centroid pairs in $ID$ in the same order}\;

\For{$pos$ $\leftarrow 1$ \KwTo $\hat{\alpha}$}{

$u, v \gets$ $ID[pos]$\;

$cluster \gets IMI_i[u, v]$  \hfill \tcp*[h]{ORAM-based read}

$R_i\gets R_i\uplus cluster$ \hfill \tcp*[h]{$\uplus$: multiset union}
}
}
{\textbf{sync}}\;
\Return $\mathcal R$\;
\caption{parallel $\FO$ candidate activation}
\label{alg:paractive}
\end{algorithm}

First, we allocate a public number $p\geq1$ of workers to each subspace, giving $w=N_s\cdot p$ workers in total. For example, $N_s=8$ and $p=10$ give $80$ workers. 
For each subspace, centroid indices are partitioned into public ranges among the $p$ workers for distance computation. These workers also form all $\kappa$ pairwise distance sums and execute parallel {\tt\small Obli\_Sort} using the standard Batcher's bitonic sorting network~\cite{DBLP:conf/afips/Batcher68}.

After the $\FO$ candidate activation in Lines~5-8, Line~9 invokes \textbf{sync} to wait for all tasks for different subspaces to complete before proceeding. Finally, Line~10 returns collection $\mathcal R=\{R_1,\ldots,R_{N_s}\}$.

\subsection{Parallel Candidate Refinement}
\label{subsec:parallel-refinement}

To parallelize $\name$'s candidate refinement, the main challenge is collision-score aggregation (Lines~1-3 of Algorithm~\ref{alg:refine}) where updates from different subspaces depend on the query and share mutable score state. To preserve the full obliviousness, directly executing these updates concurrently requires coordination of both the scores and the ORAM state, which can serialize execution. Hence, in Algorithm~\ref{alg:parrefine}, we~exploit the additive structure of collision scores to separate independent local accumulation from global aggregation in public item order. This decomposition removes cross-subspace contention during accumulation while preserving the original scoring rule.

\begin{algorithm}[tbp]
\small
\DontPrintSemicolon
\KwIn{Inputs of Algorithm~\ref{alg:refine}, and public worker count $p$}
\KwOut{Approximate top-$k$ answer set $A_k(q)$}
\For{$i\gets1$ \KwTo $N_s$ \textnormal{in parallel}}{
\For{$j\gets0$ \KwTo $n$}{
{$\mathsf{SC}_i[j]\gets0$} \tcp*[h]{ORAM-based write}\;
}
\For{$z\gets1$ \KwTo $\ell_i$ $=$ $\hat\alpha\cdot Max[i]$}{
$a\gets$ {\tt\small Obli\_Set}$(R_i[z], \bot)$\;
{$\mathsf{SC}_i[a]++$} 
\hfill \hfill \tcp*[h]{ORAM-based update}\;
}
\For{$j\gets1$ \KwTo $n$}{
{$V_i[j]\gets\mathsf{SC}_i[j]$} \hfill\hfill \tcp*[h]{ORAM-based read}\;
}
}
{\textbf{sync}}\;
\For{$j\gets1$ \KwTo $n$ \textnormal{in parallel}}{
{$\mathsf{SC}[j]\gets\sum_{i=1}^{N_s}V_i[j]$}\;
}
{\textbf{sync}}\;

invoke parallel {\tt\small Obli\_Sort} using $p$  workers to sort $\mathsf{SC}$ in descending order\;

\For{$z\gets1$ \KwTo $\beta\cdot n$}{
let $\tau$ be the item identifier for the $z^\textnormal{th}$ element in $\mathsf{SC}$\;

$W[z]\gets\langle X_\tau,\tau\rangle$ \hfill \tcp*[h]{ORAM-based read
}}

\For{$z\gets1$ \KwTo $\beta\cdot n$ \textnormal{in parallel}}{
$A[z]\gets\big\langle\rho(W[z].X,q),W[z].\tau\big\rangle$ \hfill \tcp*[h]{use $p$ workers to compute original distances}\;
}
\textbf{sync}\;

invoke parallel {\tt\small Obli\_Sort} using $p$ workers to sort the elements in $A$ in ascending order based on their distances\;

retrieve the items of the first $k$ elements in sorted $A$ through ORAM and write them into consecutive slots of $A_k(q)$\;
\Return $A_k(q)$\;

\BlankLine

\nonl \textbf{Operation} {\tt\small Obli\_Set}$(x, \bot)$:\;
\Return $x\cdot$ $\big(1$ $-$ {\tt\small Obli\_Equal}$(x, \bot)\big)$\;\nonl \tcp{\texttt{Obli\_Equal} returns $1$ iff its inputs are equal, and $0$ otherwise, without data-dependent accesses or branches}
\caption{Parallel $\FO$ candidate refinement}
\label{alg:parrefine}
\end{algorithm}

\stab
\noindent\textbf{Local Accumulation and Oblivious Aggregation.}
For each subspace $S_i$, Lines~1-6 maintain a local score array $\mathsf{SC}_i$ in an independent ORAM, with $n$ real-item slots and a reserved dummy slot at address $0$. Each subspace scans all $\ell_i=\hat{\alpha}\cdot Max[i]$ positions of its activated candidate multiset $R_i$ and performs one ORAM-protected read-modify-write update per identifier. The operation {\tt\small Obli\_Set} (Line~23) maps virtual identifiers to the dummy slot while preserving real identifiers, without data-dependent branching. Then, score accumulation proceeds concurrently across subspaces in Line~6, while accesses within each ORAM remain sequential.
After the previous local accumulation, each subspace reads its $n$ real-item scores through ORAM in public identifier order into array $V_i$, excluding the dummy slot.  Following synchronization, workers independently aggregate the scores in all subspaces according to $\mathsf{SC}[j]=\sum_{i=1}^{N_s}V_i[j], 1\leq j\leq n$.

\stab
\noindent\textbf{Parallel Selection and Distance Refinement.}
Similar to Line~4 of Algorithm~\ref{alg:refine}, Line~13 of Algorithm~\ref{alg:parrefine} uses $p$ workers and the parallel bitonic sorting procedure to rank the global scores. Lines~14-16 retrieve the first $\beta\cdot n$ items sequentially through the shared database ORAM into fixed buffer positions in array $W$. The retrieved items' original-space distances to the query are then computed in parallel using $p$ workers in Lines~17-18, stored as distance--identifier pairs in array $A$. Line~20 invokes parallel oblivious distance sorting on $A$ by the element distances in ascending order, and   
Line~21 retrieves the vectors identified by the first $k$ records in $A$ through ORAM and writes them into consecutive output slots. Finally, Line~22 returns $A_k(q)$.

\subsection{Analysis of Parallel $\name$}
\label{subsec:parallel-analysis}
\noindent\textbf{Complexity.}
We distinguish total work from parallel running time and use the black-box ORAM cost model of Section~\ref{sec:method}, accounting for the size of each accessed block.
Let $\ell=\sum_{i=1}^{N_s}\ell_i$ and let $d_i$ denote the dimension of $S_i$ such that $\sum_i d_i=d$. For Algorithm~\ref{alg:paractive}, the total work remains the same as that of sequential activation over all subspaces:
\begin{equation}
\nonumber
\label{eq:parallel-activation-work}
O\!\left(\sqrt\kappa\cdot d+N_s\cdot\kappa\cdot\log^2\kappa+\ell\cdot\log\kappa\right).
\end{equation}
A bitonic network on $n$ records has $O(n\cdot\log^2 n)$ work and $O(\log^2 n)$ depth. With $p$ workers per subspace, the time complexity of parallel activation is therefore
\begin{equation}
\nonumber
\label{eq:parallel-activation-time}
O\!\left(\max_i\left\{\left\lceil\frac{\sqrt\kappa}{p}\right\rceil\cdot d_i+\frac{\kappa\cdot\log^2\kappa}{p}+\log^2\kappa+\ell_i\cdot\log\kappa\right\}\right).
\end{equation}
As the last term includes sequential cell retrieval and copying within each subspace over its individual ORAM, these operations are not divided among its $p$ workers.

For Algorithm~\ref{alg:parrefine}, its local accumulation (Lines~4-6) performs the same $\ell$ score updates as Algorithm~\ref{alg:refine}, requiring $O(\ell\cdot\log n)$ work. Initializing and reading back the local score ORAMs (Lines~2-3 and Lines~7-8) require $O(N_s\cdot n\cdot\log n)$ work, while merging the local scores (Lines~10-11) requires $O(N_s\cdot n)$ work. The resulting total work is
\begin{equation}
\nonumber
\label{eq:parallel-refinement-work}
O\!\left((\ell+N_s\cdot n)\log n+n\cdot\log^2 n+\beta\cdot n\cdot d\cdot\log n\right).
\end{equation}
With public item ranges assigned to $p$ workers for merging, the time complexity of parallel refinement is
\begin{equation}
\nonumber
\label{eq:parallel-refinement-time}
\begin{aligned}
O\!\Big(& \big(n+\max_i\{\ell_i\}\big)\cdot\log n+N_s\cdot\lceil n/p\rceil
+n\cdot\log^2 n/p\\
&+\log^2 n+\beta\cdot n\cdot d\cdot\log n+k\cdot d\Big),
\end{aligned}
\end{equation}

For the additional space consumption, assume a linear-space ORAM implementation. Then, the local score ORAMs and readback arrays require $O(N_s\cdot n)$ space, while buffering the selected vectors requires $O(\beta\cdot n\cdot d)$ space. Parallel activation additionally requires $O(N_s\cdot\kappa)$ space for the distance sums and centroid identifier pairs. Including the $O(\ell)$ activation buffers, parallel $\name$ requires $O(\ell+N_s\cdot\kappa+N_s\cdot n+\beta\cdot n\cdot d)$ auxiliary space, excluding database and index storage. It thus trades additional storage and aggregation work for concurrent score accumulation, parallel sorting, and parallel distance computation while preserving full obliviousness.

\stab
\noindent\textbf{Correctness and Obliviousness.}
For the correctness of parallel $\name$, Algorithms~\ref{alg:paractive} and~\ref{alg:parrefine} preserve the activated candidates and aggregate the same collision scores as the search base. With the same search budgets, arithmetic, and tie-breaking rules, they therefore return the same answer, yielding the following lemma.
\begin{lemma}
\label{lemma:parallel-equivalence}
Algorithms~\ref{alg:paractive} and~\ref{alg:parrefine} return the same answer as Algorithms~\ref{alg:canactive} and~\ref{alg:refine}.
\end{lemma}

Regarding the obliviousness analysis, we follow the security assumptions in Section~\ref{sec:method}. Parallel $\name$ uses independent ORAM instances and local buffers across subspaces. Its parallel execution uses public task assignments and synchronization points, and scheduling decisions do not depend on private query values or intermediate results. Specifically, local scans have public lengths, ORAMs hide query-dependent addresses, and merging and oblivious sorting follow public access schedules. Together with oblivious local operations and a fixed public amount of work per ORAM access, these properties yield an indistinguishability analysis of the complete parallel execution, establishing Theorem~\ref{theo:parallel-fo}.

\begin{theorem}
\label{theo:parallel-fo}
The parallel $\name$ satisfies full obliviousness.
\end{theorem}

\section{Experimental Evaluation}\label{sec:experiment}
We empirically evaluate $\name$ for approximate nearest neighbor search, covering the experimental setup (Section~\ref{Exp:setting}), overall evaluation (Section~\ref{Exp:overall}), parallel evaluation (Section~\ref{Exp:parallel-evaluation}), and parameter studies (Sections~\ref{Exp:collision-parameter}--\ref{Exp:index-padding}).
The source code is publicly available at
\href{https://github.com/guozhong-li/ObliSC}{https://github.com/guozhong-li/ObliSC}.

\subsection{Experimental Setup}\label{Exp:setting}
\noindent\textbf{Hardware and Software.}
We implemented a prototype of $\name$ in C++ and conducted experiments on a server equipped with $2$TB RAM and dual $64$-core \emph{AMD EPYC} 7742 processors ($2.25$GHz), which supports \emph{TEE} via the \emph{AMD SEV} technology. All methods except parallel $\name$ used $100$ worker threads, each processing one query at a time.

\begin{table}[!t]
\centering
\caption{Dataset statistics used in the evaluation.}
\label{tab:exp-datasets}
\resizebox{\columnwidth}{!}{%
\begin{tabular}{lllrr}
\toprule
Dataset & Domain & Feature type & Cardinality & Dim. \\
\midrule
Deep1M~\cite{dataset:gqr} & Image & Deep/CNN feature & 1M & 256 \\
Gist1M~\cite{dataset:gqr} & Image & GIST descriptor & 1M & 960 \\
Sift10M~\cite{dataset:texmex} & Image & SIFT descriptor & 10M & 128 \\
Yandex Deep10M~\cite{dataset:yandex-deep} & Image & Deep embedding & 10M & 96 \\
Sift100M~\cite{dataset:texmex} & Image & SIFT descriptor & 100M & 128 \\
Glove~\cite{dataset:glove-angular} & Text & Word embedding & 1.18M & 100 \\
\bottomrule
\end{tabular}
}
\end{table}

\stab
\noindent\textbf{Datasets and Queries.}
We used six ANN datasets covering image and text retrieval.
The five image datasets are Deep1M~\cite{dataset:gqr}, Gist1M~\cite{dataset:gqr}, Sift10M~\cite{dataset:texmex}, Yandex Deep10M~\cite{dataset:yandex-deep}, and Sift100M~\cite{dataset:texmex}; the text dataset is Glove~\cite{dataset:glove-angular}.
Table~\ref{tab:exp-datasets} reports the dataset domain, feature type, cardinality, and dimensionality.
Following SuCo's experimental setting, we used $100$ queries per evaluation, with the same query set across all methods for each dataset.
For Deep1M and Glove, we used the released ground truth; for Gist1M, Sift10M, Yandex Deep10M, and Sift100M, we computed exact top-$100$ ground truth by exhaustive distance evaluation over the entire dataset. All methods use the same ground-truth file for each dataset.

\stab
\noindent\textbf{Baselines.}
We evaluate the following baselines and our proposed methods in the single-server setting:
\begin{itemize}[leftmargin=*,labelindent=0pt,labelsep=0.5em,align=left]
\item \textbf{SuCo}~\cite{wei2025suco}. The plaintext subspace-collision baseline, without query or access-pattern privacy.
\item \textbf{PACMANN}~\cite{DBLP:conf/iclr/ZhouSF25}. A secure ANN baseline combining graph search with private information retrieval.
\item \textbf{BiSON}~\cite{DBLP:conf/eurosp/DasRCG26}. An ORAM-based secure ANN baseline, evaluated under local loopback and emulated WAN conditions.
\item \textbf{$\name$-FO}. Our fully oblivious search base, using fixed-budget candidate activation and refinement.
\item \textbf{$\name$-FO-par}. The parallel version of ObliSC-FO using the same oblivious backend.
\end{itemize}

Regarding the secure ANN baseline PP-ANNS~\cite{liu2025ppanns}, although it protects both the query and the dataset, it fails to preserve access pattern privacy and does not provide open-source code. Hence, we omit a direct comparison.

\stab
\noindent\textbf{Performance Metrics.}
We report index construction time~(s), average online search time~(ms), average recall@$50$, and persistent index size~(GB). Search time covers online query processing, excluding offline index construction; index size excludes transient query workspace. Recall@$50$ measures the fraction of exact top-$50$ neighbors returned (Equation~\ref{eq:ann-recall}), averaged over the $100$ queries, with $k=50$.

\stab
\noindent\textbf{Default Parameters.}
For SuCo and PACMANN, we follow the parameter settings in their original papers~\cite{wei2025suco,DBLP:conf/iclr/ZhouSF25}.
SuCo's collision ratio $\alpha$ and refinement ratio $\beta$ control candidate activation and re-ranking, respectively.
For BiSON, we use graph degree $128$, construction/search list size $125$, eight search iterations, and beamwidth $32$, with identical search parameters under local and WAN conditions.

Following SuCo, the reference configuration for our shared subspace-collision pipeline is $\alpha=0.03$, $\beta=0.003$, $\kappa=50^2$, $N_s=8$, and $t=2$.
Here, $N_s$, $\kappa$, and $t$ denote the number of subspaces, IMI cells per subspace, and K-means iterations, respectively.
$\name$-FO and $\name$-FO-par replace the collision ratio $\alpha$ with the fixed collision parameter $\hat{\alpha}$, whose default value is $100$.
For $\name$-FO-par, the default worker count per subspace is $p=2$.

\subsection{Overall Evaluation}\label{Exp:overall}
We first evaluate the overall performance of $\name$ and the baselines on all six datasets, measuring index construction time, online search time, recall, and index size in Table~\ref{tab:main-results}.

\stab
\noindent\textbf{Index Time.}
SuCo builds the plaintext subspace-collision index, PACMANN constructs graph/PIR artifacts, and $\name$-FO additionally materializes the fixed-shape public index required by Algorithm~\ref{alg:index}.
Across all datasets, $\name$-FO incurs a moderate construction overhead over SuCo because it materializes the padded public index and fixed query buffers.
Specifically, its construction time is $1.08$--$1.65\times$ that of SuCo across the six datasets. This overhead is much smaller than PACMANN's graph/PIR construction cost: on Sift100M, $\name$-FO builds the index in $99.13$s, while PACMANN requires $14,880$s.
The shared FO/FO-par index columns retain the archived FO measurements.
FO-par adds no index construction overhead.
BiSON's indexing cost is dominated by graph construction and includes one-time ORAM initialization for both local and WAN searches.

\stab
\noindent\textbf{Online Search Time.}
The reported FO search time is higher than plaintext SuCo's on every dataset.
At the same time, $\name$-FO remains well below PACMANN's online cost on all evaluated datasets: on Sift100M, $\name$-FO takes $312.74$ms per query, compared with $1,688.6$ms for PACMANN.
FO-par reports lower search times than FO on all six datasets; on Sift100M, the time decreases to $116.18$ms.
Across six datasets, BiSON's WAN search time is $21.08$--$28.61\times$ the local value, without changing recall or persistent index size.
Both variants use the same backend and search configuration, ensuring a fair comparison of parallel performance.

\stab
\noindent\textbf{Recall@50.} $\name$-FO has lower recall than SuCo on Deep1M, Gist1M, Sift10M, and Yandex Deep10M; the largest gap is $6.3$ percentage points on Gist1M ($0.871$ versus $0.934$). On Sift100M and Glove, $\name$-FO instead achieves higher recall: $0.984$ versus $0.982$, and $0.880$ versus $0.816$, respectively.
This dataset-dependent pattern is consistent with the different activation budgets: $\name$-FO probes a fixed number of IMI cells, whereas SuCo targets a candidate count, so the number of activated items depends on bucket occupancy.
For PACMANN, it gives the lowest recall among the baselines under the recall@$50$ metric.
FO-par matches FO's reported recall on all six datasets.
BiSON reaches $0.995$ recall on Sift10M but $0.688$ on Gist1M, showing dataset-dependent quality under the same search parameter settings.

\stab
\noindent\textbf{Index Size.}
$\name$-FO's IMIs include the SuCo-side public index with redundant IMI cell elements in each IMI bucket, which are used for padding to ensure full obliviousness. Its reported index size is consistently larger than SuCo's, but smaller than PACMANN's on every benchmark. In particular, on Sift100M, $\name$-FO's index size is $9.62$GB, compared with $6.4$GB for SuCo and $15.94$GB for PACMANN.
BiSON's larger startup footprint includes encrypted graph and vector records, rather than identifier-only index overhead.

\begin{table}[!t]
\centering
\caption{Overall performance on six ANN datasets.}
\label{tab:main-results}
\setlength{\tabcolsep}{3pt}
\resizebox{\columnwidth}{!}{%
\begin{tabular}{@{}llrrrr@{}}
\toprule
Data & Method & \shortstack{Index\\Time (s)} & \shortstack{Search\\Time (ms)} & Recall@50 & \shortstack{Index\\Size (GB)} \\
\midrule
\multirow{6}{*}{Deep1M} & SuCo & 2.35 & 2.55 & 0.989 & 0.08 \\
 & PACMANN & 136.04 & 1094.98 & 0.802 & 0.53 \\
 & BiSON-local & \multirow{2}{*}{596.61} & 24.81 & \multirow{2}{*}{0.971} & \multirow{2}{*}{12.01} \\
 & BiSON-WAN & & 668.11 & & \\
\rowcolor{black!8}
 \cellcolor{white} & $\name$-FO &  & 18.57 &  &  \\
\rowcolor{black!8}
 \cellcolor{white} & $\name$-FO-par & \multirow{-2}{*}{2.65} & 7.48 & \multirow{-2}{*}{0.956} & \multirow{-2}{*}{0.44} \\
\midrule
\multirow{6}{*}{Gist1M} & SuCo & 4.29 & 3.42 & 0.934 & 0.10 \\
 & PACMANN & 816.77 & 1155.52 & 0.623 & 1.49 \\
 & BiSON-local & \multirow{2}{*}{1,468.60} & 52.19 & \multirow{2}{*}{0.688} & \multirow{2}{*}{35.87} \\
 & BiSON-WAN & & 1,228.74 & & \\
\rowcolor{black!8}
 \cellcolor{white} & $\name$-FO &  & 20.50 &  &  \\
\rowcolor{black!8}
 \cellcolor{white} & $\name$-FO-par & \multirow{-2}{*}{4.98} & 8.86 & \multirow{-2}{*}{0.871} & \multirow{-2}{*}{0.35} \\
\midrule
\multirow{6}{*}{Sift10M} & SuCo & 9.38 & 6.30 & 0.974 & 0.64 \\
 & PACMANN & 1004.36 & 1282.71 & 0.793 & 2.09 \\
 & BiSON-local & \multirow{2}{*}{2,221.36} & 33.20 & \multirow{2}{*}{0.995} & \multirow{2}{*}{67.71} \\
 & BiSON-WAN & & 699.73 & & \\
\rowcolor{black!8}
 \cellcolor{white} & $\name$-FO &  & 47.36 &  &  \\
\rowcolor{black!8}
 \cellcolor{white} & $\name$-FO-par & \multirow{-2}{*}{10.89} & 31.59 & \multirow{-2}{*}{0.957} & \multirow{-2}{*}{1.26} \\
\midrule
\multirow{6}{*}{\shortstack[l]{Yandex\\Deep10M}} & SuCo & 8.32 & 5.05 & 0.968 & 0.64 \\
 & PACMANN & 975.90 & 1254.56 & 0.792 & 1.93 \\
 & BiSON-local & \multirow{2}{*}{1,867.33} & 28.41 & \multirow{2}{*}{0.990} & \multirow{2}{*}{57.86} \\
 & BiSON-WAN & & 668.69 & & \\
\rowcolor{black!8}
 \cellcolor{white} & $\name$-FO &  & 50.32 &  &  \\
\rowcolor{black!8}
 \cellcolor{white} & $\name$-FO-par & \multirow{-2}{*}{9.53} & 43.35 & \multirow{-2}{*}{0.953} & \multirow{-2}{*}{1.22} \\
\midrule
\multirow{6}{*}{Sift100M} & SuCo & 60.17 & 83.99 & 0.982 & 6.40 \\
 & PACMANN & 14,880 & 1688.60 & 0.564 & 15.94 \\
 & BiSON-local & \multirow{2}{*}{12,012.43} & 32.45 & \multirow{2}{*}{0.938} & \multirow{2}{*}{570.25} \\
 & BiSON-WAN & & 756.86 & & \\
\rowcolor{black!8}
 \cellcolor{white} & $\name$-FO &  & 312.74 &  &  \\
\rowcolor{black!8}
 \cellcolor{white} & $\name$-FO-par & \multirow{-2}{*}{99.13} & 116.18 & \multirow{-2}{*}{0.984} & \multirow{-2}{*}{9.62} \\
\midrule
\multirow{6}{*}{Glove} & SuCo & 2.73 & 2.34 & 0.816 & 0.10 \\
 & PACMANN & 163.22 & 1091.29 & 0.604 & 0.21 \\
 & BiSON-local & \multirow{2}{*}{462.52} & 21.61 & \multirow{2}{*}{0.921} & \multirow{2}{*}{13.23} \\
 & BiSON-WAN & & 618.28 & & \\
\rowcolor{black!8}
 \cellcolor{white} & $\name$-$\FO$ &  & 25.63 &  &  \\
\rowcolor{black!8}
 \cellcolor{white} & $\name$-FO-par & \multirow{-2}{*}{2.95} & 8.42 & \multirow{-2}{*}{0.880} & \multirow{-2}{*}{0.17} \\
\bottomrule
\end{tabular}
}
\end{table}

\subsection{Parallel Evaluation}\label{Exp:parallel-evaluation}
We examined how workers per subspace $p$ affect $\name$-FO-par's search efficiency by varying $p$ from $1$ to $5$ on all six datasets, with other parameters unchanged (shown in Table~\ref{tab:parallel-worker-count}). The lowest average batch search time per query occurs at $p=2$ on every dataset. Increasing $p$ from $1$ to $2$ reduces the time from $15.77$ms to $7.48$ms on Deep1M and from $178.43$ms to $116.18$ms on Sift100M. Across datasets, the speedup over $p=1$ ranges from $1.17\times$ on Yandex Deep10M to $2.34\times$ on Glove. Further increasing $p$ generally worsens performance: at $p=5$, the corresponding times rise to $36.02$ms and $323.29$ms.

\begin{table}[!t]
\centering
\caption{Effect of $p$ on search time (ms).}
\label{tab:parallel-worker-count}
\setlength{\tabcolsep}{4pt}
\small
\begin{tabular}{@{}lrrrrr@{}}
\toprule
Dataset & $p=1$ & $p=2$ & $p=3$ & $p=4$ & $p=5$ \\
\midrule
Deep1M & 15.77 & \textbf{7.48} & 20.61 & 27.21 & 36.02 \\
Gist1M & 18.63 & \textbf{8.86} & 20.81 & 25.26 & 32.51 \\
Sift10M & 62.15 & \textbf{31.59} & 74.10 & 84.59 & 111.09 \\
Yandex Deep10M & 50.72 & \textbf{43.35} & 62.28 & 70.64 & 108.80 \\
Sift100M & 178.43 & \textbf{116.18} & 249.40 & 206.42 & 323.29 \\
Glove & 19.67 & \textbf{8.42} & 25.08 & 31.39 & 37.49 \\
\bottomrule
\end{tabular}
\end{table}

This trend suggests a tradeoff between parallel speedup and execution overhead. Although more workers can accelerate computation, parallel oblivious sorting requires synchronization between dependent network layers, and additional workers may incur greater coordination and memory-contention costs. Beyond $p=2$, these overheads appear to outweigh the gains from further parallelism. The observed performance supports our default choice of $p=2$.

\subsection{Collision-Parameter Tradeoff}\label{Exp:collision-parameter}
\begin{figure}[!t]
\centering
\includegraphics[width=\columnwidth]{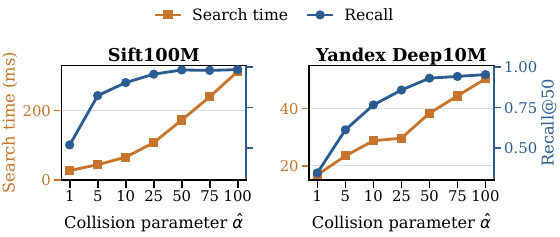}
\caption{Effect of $\hat{\alpha}$ on search time and recall.} %
\label{fig:collision-parameter-tradeoff}
\Description{Two side-by-side line plots for Sift100M and Yandex Deep10M. Search time and recall generally increase with the collision parameter, while recall approaches saturation at larger values.}
\end{figure}

We evaluated how the collision parameter $\hat{\alpha}$ affects $\name$'s recall and search time, since $\name$'s candidate activation (Algorithm~\ref{alg:canactive}) accesses exactly $\hat{\alpha}$ IMI cells per subspace. We varied $\hat{\alpha}$ over $1$, $5$, $10$, $25$, $50$, $75$, and $100$ on Sift100M and Yandex Deep10M and reported the results in Figure~\ref{fig:collision-parameter-tradeoff}.

It can be observed from Figure~\ref{fig:collision-parameter-tradeoff} that $\name$ exhibits the expected quality-cost tradeoff on both datasets:
on Sift100M, $\name$'s recall rises from $0.518$ at $\hat{\alpha}=1$ to $0.981$ at $\hat{\alpha}=50$ and then saturates near $0.984$;
on Yandex Deep10M, a similar pattern appears, but the recall benefits more gradually:
it increases from $0.346$ to $0.931$ by $\hat{\alpha}=50$ and reaches $0.953$ at $\hat{\alpha}=100$.
Meanwhile, the search time on both datasets grows as $\hat{\alpha}$ increases. The reason for such trends is that a larger $\hat{\alpha}$ leads to more IMI cell accesses, taking $\name$ more time to activate more candidates.

\subsection{Query Parameter Sensitivity}\label{Exp:topk-sensitivity}
\begin{figure}[!t]
\centering
\includegraphics[width=\columnwidth]{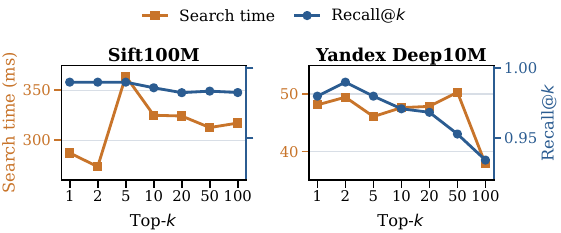}
\caption{Search-time and recall sensitivity to top-$k$.}
\label{fig:topk-sensitivity}
\Description{Two side-by-side dual-axis plots showing absolute search time and recall as top-k changes from 1 to 100 on Sift100M and Yandex Deep10M.}
\end{figure}

We examined how the query parameter $k$ affects $\name$'s search efficiency by varying $k$ over $1$, $2$, $5$, $10$, $20$, $50$, and $100$ on Sift100M and Yandex Deep10M, and reported the results in Figure~\ref{fig:topk-sensitivity}.

We can observe from Figure~\ref{fig:topk-sensitivity} that Recall@$k$ remains high across all $k$:
$\name$'s recall on Sift100M stays between $0.983$ and $0.99$, while the recall on Yandex Deep10M changes smoothly from $0.98$ at $k=1$ to $0.934$ at $k=100$. However, on both datasets, the recall decreases as $k$ increases. This is because $k$ affects only the final truncation in $\name$'s candidate refinement, making it harder to recover a larger set of exact nearest neighbors from an unchanged set of candidates. Regarding the search efficiency, we can see that $\name$'s search time on Sift100M remains around $300$ms for all $k$, with reasonable fluctuation across runs. A similar trend can also be observed for Yandex Deep10M.

\subsection{Refinement Budget Sensitivity}\label{Exp:budget-sensitivity}
\begin{figure}[!t]
\centering
\includegraphics[width=\columnwidth]{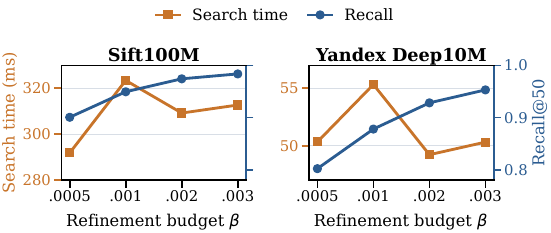}
\caption{Search-time and recall sensitivity to $\beta$.}
\label{fig:sc-sensitivity}
\Description{Two side-by-side dual-axis plots showing absolute search time and recall as beta changes on Sift100M and Yandex Deep10M.}
\end{figure}

The refinement budget, namely the re-rank ratio $\beta$, determines that $\name$'s candidate refinement (Algorithm~\ref{alg:refine}) must refine  $\beta \cdot n$ candidates. To investigate how $\beta$ affects $\name$'s performance, we varied $\beta$ over $0.0005$, $0.001$, $0.002$, and $0.003$ on Sift100M and Yandex Deep10M. The experimental results are reported in Figure~\ref{fig:sc-sensitivity}.

As shown in Figure~\ref{fig:sc-sensitivity}, increasing $\beta$ consistently improves the utility of $\name$'s output result:
recall rises from $0.901$ to $0.984$ on Sift100M and from $0.802$ to $0.953$ on Yandex Deep10M.
However, $\name$'s search time remains within a comparatively narrow band because $\name$'s candidate activation dominates these runs, although a larger $\beta$ takes $\name$  more time for computing more original distances and sorting more elements.
Fluctuations of $25$ms on a $300$ms search time remain within an acceptable range of experimental variance.
For Yandex Deep10M, the larger gain in recall when increasing $\beta$ indicates greater benefit from re-ranking additional collided candidates in $\mathsf{SC}$; 
the search times follow a similar trend to those on Sift100M.
Thus, $\beta$ is principally a recall knob once the collided candidates in $\mathsf{SC}$ have been fixed, and the default $\beta=0.003$ favors the high-recall end of this tradeoff.

\subsection{K-means Parameter Sensitivity}\label{Exp:index-padding}
We further studied how the K-means parameters, namely $\kappa$ and $t$, govern IMI bucket balance.

\begin{figure}[!t]
\centering
\includegraphics[width=\columnwidth,trim=0 5bp 0 0,clip]{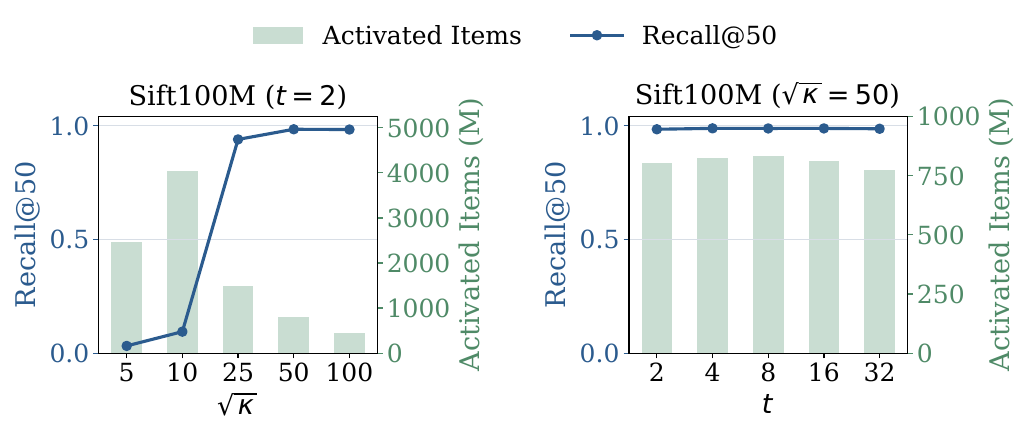}
\caption{Sensitivity to $\sqrt{\kappa}$ and $t$ on Sift100M.}
\label{fig:kmeans-sensitivity}
\Description{Two side-by-side plots showing activated items and recall under square-root-kappa and t sweeps.}
\end{figure}

\stab
\noindent\textbf{IMI Granularity $\sqrt{\kappa}$.}
{The IMI granularity $\sqrt{\kappa}$ is the number of K-means clusters in each half-subspace, giving $\kappa$ cells per IMI.}
We first varied $\sqrt{\kappa}$ over $5$, $10$, $25$, $50$, and $100$, and investigated how $\sqrt{\kappa}$ affects the recall and the number of activated items (including padding across subspaces) in $\name$'s candidate activation (Algorithm~\ref{alg:canactive}) on Sift100M.
As shown by the LHS subfigure of Figure~\ref{fig:kmeans-sensitivity}, moderate granularity balances recall and activated items: recall rises from $0.032$ at $\sqrt{\kappa}=5$ to $0.984$ at $\sqrt{\kappa}=50$, while finer cells reduce activation.
Activated items peak at $\sqrt{\kappa}=10$ and then decrease. However, at $\sqrt{\kappa}=100$, fewer items are activated but recall drops slightly.
This is because overly fine granularity may cause some true-positive items to be excluded from the activated IMI bucket cell.

\stab
\noindent\textbf{K-means Clustering Iterations $t$.}
We varied the K-means clustering iterations $t$ over $2$, $4$, $8$, $16$, and $32$, and also reported the recall and the number of activated items in the RHS subfigure of Figure~\ref{fig:kmeans-sensitivity}.
Activated items first increase from $804.50$M at $t=2$ to $833.18$M at $t=8$, then decrease to $771.67$M at $t=32$.
Recall remains between $0.984$ and $0.988$, with no monotonic gain from more iterations, supporting $t=2$ as the default.

\section{Related Work}
\label{sec:related}

\noindent\textbf{Index-Based ANN Search.} 
Common ANN indexing and search techniques include product quantization~\cite{jegou2011product}, locality-sensitive
hashing (LSH)~\cite{indyk1998approximate}, inverted files, trees~\cite{li2025lesax}, and graph-based methods such as HNSW~\cite{malkov2020hnsw}.
Recently, \emph{subspace collision} (SC)~\cite{wei2025suco} has offered another approach:
it aggregates collision information across multiple lower-dimensional subspaces to estimate proximity in the original space.
Building on this framework, SuCo~\cite{wei2025suco} combines a lightweight clustering-based index with
efficient query-processing strategies.

The efficiency gains of these methods often rely on
query-dependent behavior, including bucket and cluster
selection, graph traversal, pruning, and candidate
refinement.
While such adaptivity is natural in plaintext retrieval
systems, the resulting access patterns can leak
information about a private query vector.
Designing ANN search methods that protect query privacy
against access-pattern leakage while maintaining high
search efficiency therefore remains an important problem.

\stab
\noindent\textbf{Privacy-Preserving ANN Search.}
In the single-server setting, Liu et al.~\cite{liu2025ppanns} studied ANN search with protection for both query and database contents, but their construction does not hide query-dependent access patterns. Li et al.~\cite{DBLP:conf/ccs/LiHZ0L0C25} proposed Panther, which combines PIR, secret sharing, garbled circuits, and homomorphic encryption to protect both query and database privacy while hiding query-dependent access patterns against semi-honest adversaries. Das et al.~\cite{DBLP:conf/eurosp/DasRCG26} proposed BiSON, which combines DiskANN with a disk-compatible ORAM architecture to protect database contents, queries, and access patterns from an untrusted server in an outsourced setting. Zhou et al.~\cite{DBLP:conf/iclr/ZhouSF25} proposed Pacmann, a graph-based ANN search framework that protects query privacy and hides query-dependent graph accesses via Private Information Retrieval (PIR) over a public database. Like Pacmann, $\name$ protects queries over a public vector database, using a trusted execution environment (TEE) for query confidentiality and oblivious search within the TEE to hide access patterns.

In the two-server setting, Servan-Schreiber et al.~\cite{DBLP:conf/sp/Servan-Schreiber22} proposed Preco, a locality-sensitive hashing-based private ANN protocol using distributed point functions for private retrieval from two non-colluding servers, protecting the query while limiting database information disclosed to the client. In the federated setting, Zhang et al.~\cite{DBLP:journals/pacmmod/ZhangWXPX24} proposed FedKNN for secure kNN search across private data providers. FedKNN combines TEE with differential obliviousness to protect the providers' data against access-pattern leakage.

\stab
\noindent{\bf Oblivious Search.} 
Full obliviousness~\cite{DBLP:conf/stoc/Goldreich87} provides strong privacy guarantees by making observable access patterns indistinguishable across inputs of the same public size.
Prior work has employed oblivious designs, under different security models and leakage guarantees, for a wide range of query and data-processing tasks, including SQL queries~\cite{DBLP:conf/icdt/ArasuK14,
DBLP:journals/pvldb/BaterEEGKR17,
DBLP:journals/pvldb/EskandarianZ19,
DBLP:journals/pvldb/KrastnikovKS20,
DBLP:conf/uss/MavrogiannakisW25, DBLP:journals/pacmmod/WuDH26},
graph algorithms such as traversal, shortest-path
computation, and minimum spanning
trees~\cite{DBLP:conf/ccs/BlantonSA13,
DBLP:conf/sp/NayakWIWTS15,
DBLP:journals/pvldb/ChamaniD0PJ23},
and graph pattern queries~\cite{DBLP:conf/icde/FanCXB15,
xu2021strongsimulation, xu2023localizedgraph,
DBLP:journals/tifs/WangZJW24, DBLP:journals/tkde/XuHZJYCX26}.

\section{Conclusion}\label{sec:conclusion}
In this paper, we study asymmetric privacy-preserving ANN search over a public vector database, where user queries must remain private from an untrusted cloud server. We propose $\name$, an efficient search framework whose ANN index, candidate activation, and candidate refinement are jointly designed to protect query privacy against access-pattern leakage. We prove that its search base satisfies the privacy guarantee of full obliviousness. We further develop parallel strategies for candidate activation and refinement while preserving full obliviousness and the search base's results. Experiments on six standard ANN benchmarks demonstrate that $\name$ achieves competitive efficiency and result quality while maintaining a lightweight index compared with the evaluated privacy-preserving approaches.
As for future work, we may optimize each of~$\name$'s components, implement $\name$ using alternative cryptographic tools rather than a TEE to protect query values and private intermediate values, and extend $\name$ to support additional query workloads.

\clearpage

\flushbottom
\section*{AI/LLM Usage Disclosure}
Generative AI tools were used only to polish the writing and assist with code implementation.
All technical contributions, including the proposed method, analysis, and experiments, were developed and verified by the authors.

\bibliographystyle{IEEEtran}
\bibliography{main}

\clearpage
\appendices

\balance
\section{Proofs}
\label{app:proofs}
Throughout the proofs, we use the privacy assumptions stated in Section~\ref{sec:method}.

\vspace{2pt}

\stab
\noindent\textbf{Lemma~\ref{lemma:two}.}
\emph{{\tt\small Top$k$Refine} satisfies full obliviousness.}
\begin{proof}
Consider two executions with the same public inputs but different queries and activated identifiers. By Lemma~\ref{lemma:base}, each $R_i$ has public size $\hat\alpha\cdot Max[i]$. Algorithm~\ref{alg:refine} scans these arrays in fixed order, performing one ORAM-protected read-modify-write per identifier, including virtual identifiers through the dummy slot. It then obliviously sorts the $n$ real-item scores, retrieves exactly $\beta\cdot n$ vectors through ORAM, and computes their distances using fixed local access schedules. 
The distance sort is oblivious, and the final $k$ vectors are retrieved through ORAM and written to public output positions while their contents remain hidden from SP.
Thus, local schedules and request counts are query-independent. Adaptive ORAM security hides the private addresses throughout these operations, yielding computationally indistinguishable traces and establishing Equation~\eqref{eqa:fo}.
\end{proof}

\noindent\textbf{Theorem~\ref{theo:one}.}
\emph{$\name$ satisfies full obliviousness.}
\begin{proof}
Consider two executions of $\name$ with the same public database, index, and parameters but different queries. Algorithm~\ref{alg:index} constructs the index independently of the query. Algorithm~\ref{alg:ppsc} then performs initialization and query projection using access schedules determined by public sizes and the index partition. It invokes candidate activation and refinement in a fixed order. By Lemmas~\ref{lemma:base} and~\ref{lemma:two}, both stages have query-independent local schedules and ORAM request counts. Adaptive ORAM security hides the private addresses across the complete sequence, including requests sharing ORAM state, while private intermediate values and results remain hidden from SP. Thus, sequentially composing these stages yields computationally indistinguishable traces and establishes Equation~\eqref{eqa:fo}.
\end{proof}

\noindent\textbf{Lemma~\ref{lemma:parallel-equivalence}.}
\emph{Algorithms~\ref{alg:paractive} and~\ref{alg:parrefine} return the same answer as Algorithms~\ref{alg:canactive} and~\ref{alg:refine}.}
\begin{proof}
Consider the sequential and parallel executions with the same index, query, budgets, arithmetic, and tie-breaking rules. Algorithms~\ref{alg:canactive} and~\ref{alg:paractive} compute and rank the same centroid-pair distances and retrieve the same first $\hat\alpha$ padded cells. Thus, each $R_i$ contains the same identifiers with the same multiplicities. Algorithm~\ref{alg:parrefine} counts each real identifier's occurrences locally and sums these counts, reproducing the global scores in Algorithm~\ref{alg:refine}. Virtual identifiers update only dummy slots, and counters accommodate the public maximum without overflow. Both procedures therefore select the same $\beta\cdot n$ identifiers. The distance computations use the same arithmetic order, and both sorts use the same tie-breaking rules. Consequently, the final top-$k$ answers are identical.
\end{proof}

\noindent\textbf{Theorem~\ref{theo:parallel-fo}.}
\emph{Parallel $\name$ satisfies full obliviousness.}
\begin{proof}
Consider two parallel executions with the same public inputs and worker configuration but different queries. Under the scheduling conditions in Section~\ref{sec:parallel}, Algorithm~\ref{alg:paractive} assigns distance computations, sorting comparisons, and padded-cell reads to workers independently of the query. Each output $R_i$ has public size $\ell_i$. Algorithm~\ref{alg:parrefine} initializes $n+1$ score slots per subspace, performs exactly $\ell_i$ updates including dummy updates, and reads back $n$ real scores. The ORAM instances have independent randomness and states. Applying adaptive ORAM security to each instance in turn yields indistinguishable joint traces, even though their requests may be correlated through the query.

Merging and sorting follow public access schedules. Candidate retrieval and final result retrieval issue exactly $\beta\cdot n$ and $k$ requests through the shared database ORAM, respectively. Distance computations use public buffer positions and worker assignments, and output writes visit fixed positions while their contents remain hidden from SP. Oblivious local operations and public per-access work bounds prevent additional value-dependent leakage. Under the conditions in Section~\ref{sec:parallel}, scheduling and synchronization introduce no private dependence beyond the protected traces. Thus, the complete observable executions, including thread identities and synchronization events, are computationally indistinguishable, establishing Equation~\eqref{eqa:fo} up to the negligible error of the secure primitives.
\end{proof}

\end{document}